\documentclass{aa}
\usepackage{txfonts}
\usepackage{graphicx}
\usepackage{t1enc}
\usepackage{epsfig}
\usepackage{rotating}
\usepackage{verbatim}
\usepackage[authoryear]{natbib}
\bibpunct{(}{)}{;}{a}{}{,}
\newcommand{\Td}    {T_\mathrm{d}}

\newcommand{\mum}   {$\mu$m}
\newcommand{\kms}   {km~s$^{-1}$}
\newcommand{\cmg}   {cm$^{2}$~g$^{-1}$}

\newcommand{\jpb}   {$\rm Jy~beam^{-1}$}    
\newcommand{\lo}    {$L_{\sun}$}
\newcommand{\mo}    {$M_{\sun}$}
\newcommand{\nh}    {NH$_3$}

\newcommand{\chtoh} {CH$_3$OH}
\newcommand{\water} {H$_2$O}

\newcommand{\et}    {et al.}
\newcommand{\eg}    {e.\,g.,}
\newcommand{\hii}   {\ion{H}{ii}}
\newcommand{\uchii} {UC\ion{H}{ii}}

\newcommand{\phn}   {\phantom{0}}
\newcommand{\phnn}  {\phantom{0}\phantom{0}}

\newcommand{\php}   {\phantom{.}}
\newcommand{\phpm}  {\phantom{$-$}}
\newcommand{\phs}   {\phantom{$^0$}}
\newcommand{\phe}   {\phantom{$^\mathrm{c}$}}
\newcommand{\phm}   {\phantom{$>.$}}
\newcommand{\phb}   {\phantom{$>$}}
\newcommand{\phnm}  {\phantom{0}\phantom{$>.$}}

\newcommand{\raun}  {$^\mathrm{h~m~s}$}
\newcommand{\deun}  {$\mathrm{\degr~\arcmin~\arcsec}$}
\begin{document}

   \title{Survey of intermediate/high-mass star-forming regions\\
   at centimeter and millimeter wavelengths}

%%%   \title{Ionized gas from intermediate/high-mass\\
%%%   YSOs and the associated dust}

%%%   \title{Survey of centimeter and millimeter emission\\ from high-mass
%%%   star-forming regions}

   \author{\'Alvaro S\'anchez-Monge\inst{1}
   	  \and
   	  Aina Palau\inst{1,2}
          \and
          Robert Estalella\inst{1}
	  \and
	  Maria T. Beltr\'an\inst{1}
	  \and
	  Josep M. Girart\inst{3}
          }

   \offprints{\'Alvaro S\'anchez-Monge,\\ \email{asanchez@am.ub.es}}

   \institute{Departament d'Astronomia i Meteorologia, Universitat de Barcelona,
              Mart\'i i Franqu\`es 1, E-08028 Barcelona, Spain
      	      \and
	      Laboratorio de Astrof\'{\i}sica Espacial y F\'{\i}sica
	      Fundamental, INTA, Apartado 78, E-28691 Villanueva de la Ca\~nada,
	      Madrid, Spain
	      \and
	      Institut de Ci\`encies de l'Espai (CSIC-IEEC), Campus UAB --
	      Facultat de Ci\`encies, Torre C5 -- parell 2, E-08193 Bellaterra,
	      Catalunya, Spain
      	      }

%	      Laboratorio de Astrof\'{\i}sica Espacial y F\'{\i}sica Fundamental,
%	      INTA, Apartado 50727, E-28080 Madrid, Spain
              
   \date{Received / Accepted}

   \authorrunning{\'A. S\'anchez-Monge \et}
%%%   \titlerunning{Ionized gas and dust emission from massive YSOs
%%%		}
   \titlerunning{Survey of cm and mm emission from massive star-forming regions
		}

% \abstract{}{}{}{}{} 
% 5 {} token are mandatory

\abstract{}{The goal of this work is to characterize the millimeter
and centimeter properties of intermediate/high-mass young stellar
objects (YSOs) in order to search for any evolutionary trend.}{We
carried out observations at 1.2~mm with the IRAM~30\,m telescope, and
at 3.6 and 1.3~cm with the VLA toward a sample of 11 luminous
($>10^3$~\lo) IRAS sources classified as high-mass protostellar object
candidates. The most promising regions were additionally observed at
7~mm with the VLA.}{The 1.2~mm emission is detected toward all the
IRAS sources. In some cases the emission shows a clear peak surrounded
by some substructure, while in other cases the emission is very
extended and weak. The 1.2~mm emission is likely tracing the dust core
in which the intermediate/high-mass YSO is forming, for which we
estimated masses ranging from 10 to 140~\mo. For all (but one of) the
sources, we detected centimeter emission associated with the IRAS
source, being compact or ultra-compact (UC) \hii\ region
candidates. From the spectral index between 3.6 and 1.3~cm, we
identified the possible emission mechanism in the centimeter
range. Deconvolved sizes of the centimeter sources range from $<0.01$
to 0.3~pc, and the physical parameters of the \uchii\ regions indicate
that the ionizing stars are early B-type stars. The 7~mm emission is
partially resolved for the four regions observed at this wavelength,
and we estimated the contribution of the dust emission to the 7~mm
flux density, ranging from negligible to 44\%. By combining our data
with the infrared surveys 2MASS, MSX and IRAS, and
IRAC-\emph{Spitzer} data (when available), we built the spectral
energy distribution and fitted a modified blackbody law taking into
account the contribution of the free-free centimeter emission. We
found dust temperatures between 25 and 35~K, dust emissivity indices
between 1.5 and 2.2, and masses similar to the masses derived from the
1.2~mm continuum emission. In addition, we found a correlation between
the degree of disruption of the natal cloud, estimated from the
fraction of dust emission associated with the centimeter source
relative to the total amount of dust in its surroundings, and the size
of the centimeter source.}{From the correlation found, we
established an evolutionary sequence in which sources with compact
millimeter emission clearly associated with compact centimeter
emission are younger than sources with the millimeter emission
dispersed and with the centimeter emission extended. Such a sequence
is consistent with the evolutionary stage expected from
maser/outflow/dense gas emission reported in the literature, and with
the infrared excess of the 2MASS sources associated with the
centimeter source, estimated in this work.}


\maketitle

\section{Introduction}

High-mass stars ($M\gtrsim10$~\mo) play a major role in the physical
and chemical evolution of their host galaxies, through strong
ultraviolet radiation and winds, and through supernova explosions.
However, the formation scenario of these stars is not well understood,
due to the large distances ($\gtrsim1$~kpc) and heavy extinctions
($A_{\mathrm{v}}\gtrsim100$) of the regions where they form. This,
coupled with the fact that high-mass stars seem to form in clustered
mode
\citep{testi1998,testi2000,hillenbrandphd,hillenbrandhartmann1998},
where a single massive dense core of gas forms a group of stars, makes
observationally more difficult to study these regions. A result
recently found in massive star formation is that outflows, some of
them collimated \citep[\eg][]{beuther2002,gibb2003,zhang2005},
disk-like structures
\citep[\eg][]{cesaroni1997,cesaroni1999,patel2005}, and infall motions
\citep[\eg][]{beltran2006a,zapata2008} have been detected in massive
protostars, suggesting that there is disk-mediated accretion in
high-mass stars, and that they may form as a scaled-up version of the
low-mass stars. However, the different evolutionary stages describing
how the massive protostar approaches the main-sequence may differ from
those for low-mass stars, since massive stars evolve much faster to
the main-sequence \citep[\eg][]{bernasconimaeder1996}, and, as a
consequence, the massive protostar produces UV photons while is still
accreting matter. Thus, the centimeter emission, tracing the ionized
gas, and the millimeter emission, usually tracing the dust, are
crucial to understand the first stages in the formation of a massive
star.

The work presented here is part of a large project aimed at studying
in detail (with high sensitivity and high angular resolution) a sample
of high-mass star-forming regions ($L_\mathrm{bol}\gtrsim1000~$~\lo)
in different evolutionary stages, in order to characterize the
millimeter and centimeter properties as the massive young stellar
object (YSO) evolves. For this we selected, from the
\citet{molinari1996}, \citet{mueller2002}, and \citet{sridharan2002}
samples, regions located at distances $\lesssim3$~kpc, in order to be
sensitive down to masses $\lesssim\!1$~\mo\ (from the millimeter
emission), and to have a spatial resolution $\lesssim\!5000$~AU. For
each region, we searched the literature for centimeter and millimeter
emission, and found 11 regions with no imaging at centimeter or
millimeter observations.  To complete the sample of nearby and
luminous massive star-forming regions observed at millimeter and
centimeter wavelengths, we carried out VLA and IRAM~30\,m observations
toward these 11 regions. Nine regions were observed in the millimeter
range, and 10 in the centimeter range (8 coinciding with the regions
observed at millimeter wavelengths, see Table~\ref{tregions}).

In \S~\ref{sobs}, we summarize the observations, in \S~\ref{sres} we
give the general results for the millimeter and centimeter data, in
\S~\ref{sdisc} we discuss the nature of the centimeter emission,
comment the results for each individual source, and discuss global
properties possibly indicating different evolutionary stages of the
massive YSOs. Finally, in \S~\ref{scon} we summarize our main
conclusions.

%-------------------------------------------------------------------------------
\begin{table}
\caption{List of regions observed in this work}
\begin{center}
\begin{tabular}{lcccc}
\noalign{\smallskip}
\hline\hline\noalign{\smallskip}
&$L_\mathrm{bol}$
&Dist.
\\
Region
&(\lo)
&(kpc)
&Instrument$^\mathrm{a}$
&Ref.
\\
\noalign{\smallskip}
\hline\noalign{\smallskip}
IRAS~00117+6412			&\phn1300	&1.8	& V + I		& 1\\
IRAS~04579+4703			&\phn4000	&2.5	& V + I		& 1\\
IRAS~18123$-$1203		&\phn7900	&3.1	& V + I		& 1\\
IRAS~18171$-$1548$^\mathrm{b}$	&\phn7900	&2.6	& V + I		& 1\\
IRAS~18212$-$1320		&\phn1600	&2.2	& I		& 1\\
IRAS~19045+0813			&\phn1600	&1.6	& V + I		& 1\\
IRAS~20293+4007			&\phn7900	&3.4	& V + I		& 1\\
IRAS~22187+5559			&\phn7900	&2.9	& V + I		& 1\\
IRAS~22198+6336			&\phn1300	&1.3	& V		& 1\\
NGC\,7538-IRS9$^\mathrm{c}$	&40000		&2.8	& V		& 2\\
IRAS~23448+6010			&\phn2500	&2.0	& V + I		& 1\\
\hline
\end{tabular}
\begin{list}{}{}
\item[$^\mathrm{a}$] V: VLA, I: IRAM~30\,m.
\item[$^\mathrm{b}$] This source appears as IRAS~18172$-$1548 in \citet{molinari1996}.
\item[$^\mathrm{c}$] Identified with IRAS~23118+6110.
\item[References:] 
(1) \citealt{molinari1996}, 
(2) \citealt{mueller2002}.
\end{list}
\end{center}
\label{tregions}
\end{table}
%-------------------------------------------------------------------------------

%%%%%%%%%%%%%%%%%%%%%%%%%%%%%%%%%%%%%%%%%%%%%%%%%%%%%%%%%%%%%%%%%%%%%%%%%%%%%%%%
%
% OBSERVATIONS

\section{Observations \label{sobs}}

The MPIfR 117-element bolometer array MAMBO-2 at the IRAM~30\,m
telescope\footnote{IRAM is supported by INSU/CNRS (France), MPG
(Germany), and IGN (Spain).} was used to map the 1.2~mm dust continuum
emission toward the 9 selected high-mass protostar candidates (see
Table~\ref{tregions}). The observations were carried out on 2004
January 23 to 25. The main beam of the IRAM\,30\,m at 1.2~mm has a
HPBW size of $11''$.  We used the on-the-fly mapping mode, in which
the telescope scans in azimuth along each row in such a way that all
pixels see the source once. The sampled area was typically $320''
\times 80''$, and the scanning speed was 6$''$~s$^{-1}$. Each scan was
separated by $8''$ in elevation. The secondary mirror was wobbling at
a rate of 2~Hz in azimuth with a wobbler throw of $70''$. The average
zenith opacity at 225~GHz was 0.3--0.4. The overall uncertainty in
calibration is estimated to be 20\%.  Data reduction was performed
with the MOPSIC software package (distributed by R.  Zylka).

The centimeter observations were carried out on 2004 July 10 and 13
using the Very Large Array (VLA) of the NRAO\footnote{The National
Radio Astronomy Observatory is a facility of the National Science
Foundation operated under cooperative agreement by Associated
Universities, Inc.} in the D configuration. The regions observed at
3.6~cm, the phase calibrators used and their bootstrapped fluxes, as
well as the parameters of the synthesized beam, the robust parameter
of \citet{briggs1995}, and the rms noise of the maps are given in
Table~\ref{t3cmobs}, and the same information for 1.3~cm is given in
Table~\ref{t1cmobs}. The positions of the phase centers correspond to
the coordinates given in \citet{molinari1996} or \citet{mueller2002}.
In all cases, except for IRAS~19045+0813 and NGC\,7538-IRS9, they
correspond to the catalog positions of the IRAS sources. The
integration time for each source was about 5 minutes at 3.6~cm and 11
minutes at 1.3~cm. Absolute flux calibration was achieved by observing
3C48, with an adopted flux density of 3.15~Jy and 1.12~Jy at 3.6 and
1.3~cm, and 3C286, with 5.21~Jy and 2.52~Jy at 3.6 and 1.3~cm
respectively. Images were constructed using a variety of weighting
schemes by varying the Briggs robust parameter from +1 to +5 (roughly
equivalent to natural weight) for most of the regions (see
Table~\ref{t3cmobs} and \ref{t1cmobs}). In order to increase the SNR
of the faint sources at 1.3~cm, we tapered the $uv$-data at
50~k$\lambda$, except for IRAS~22187+5559, for which we tapered the
$uv$-data at 25~k$\lambda$, in order to recover the extended emission.

The 7~mm observations carried out toward four regions (see
Table~\ref{t7mmobs}) were made with the VLA in the D configuration
during the first months of 2007, with 8 EVLA antennas in the array. In
order to minimize the effects of atmospheric phase fluctuations, we
used the technique of \emph{fast switching}
\citep{carilliHoldaway1997} between the source and the phase
calibrator over a cycle of 120 seconds, with 80 seconds spent on the
target and 40 seconds on the calibrator. The integration time for each
source was about 50 minutes, except for IRAS~04579+4703 with an
on-source time of about 100 minutes. The images were constructed using
natural weighting and tapering the $uv$-data at 60~k$\lambda$, except
for IRAS~22198+6336, to increase the SNR.

For simplicity, we will refer to each IRAS source by an "I" followed
by the RA identification of the catalog name.

%-------------------------------------------------------------------------------
\begin{table*}
\caption{Parameters of the observations at 3.6~cm with the VLA}
\begin{center}
\begin{tabular}{lcccccccc}
\noalign{\smallskip}
\hline\hline\noalign{\smallskip}
&\multicolumn{2}{c}{Phase Center}
&Phase
&Bootstrapped Flux
&Beam
&P.A.
&
&rms
\\
\cline{2-3}\noalign{\smallskip}
Region
&$\alpha$(J2000.0)
&$\delta$(J2000.0)
&Calibrator
&(Jy)		
&(arcsec)
&($\degr$)
&Robust
&($\mu$Jy~beam$^{-1}$)
\\
\noalign{\smallskip}
\hline\noalign{\smallskip}
IRAS~00117+6412
	&00 14 27.73  &+64 28 46.2    &0102+584	  &$1.885\pm0.003$   &$14.9\times8.5$		&\phpm76	&+1	&\phnn38\phe\\
IRAS~04579+4703
	&05 01 39.73  &+47 07 23.1    &0423+418	  &$1.986\pm0.004$   &$10.7\times9.4$		&\phpm48	&+5	&\phnn22\phe\\
IRAS~18123$-$1203
	&18 15 07.33  &$-$12 02 42.0  &1822$-$096 &$1.364\pm0.002$   &$12.3\times8.1$		&\phpm\phn5	&+1	&\phnn43\phe\\
IRAS~18171$-$1548
	&18 20 04.62  &$-$15 46 46.6  &1822$-$096 &$1.364\pm0.002$   &$13.0\times8.1$		&\phpm\phn2	&+1	&\phn185\phe\\
IRAS~19045+0813
	&19 06 59.77  &+08 18 42.8    &1922+155	  &$0.666\pm0.002$   &$\phn8.9\times7.8$	&\phpm43	&$-$3	&\phnn54\phe\\
IRAS~20293+4007
	&20 31 07.92  &+40 17 22.8    &2025+337	  &$1.826\pm0.003$   &$\phn7.5\times6.5$	&$-$37		&$-$5	&\phnn58\phe\\
IRAS~22187+5559
	&22 20 34.89  &+56 14 39.5    &2148+611	  &$0.971\pm0.003$   &$\phn7.8\times6.6$	&\phn$-$1	&$-$5	&\phn155\phe\\
IRAS~22198+6336
	&22 21 27.62  &+63 51 42.2    &2148+611	  &$0.971\pm0.003$   &$10.4\times8.8$		&\phpm\phn7	&+5	&\phnn27\phe\\
NGC\,7538-IRS9
	&23 14 01.68  &+61 27 19.9&0014+612	  &$1.343\pm0.004$   &$10.2\times9.3$		&$-$14		&+5	&\phn700$^\mathrm{a}$\\
IRAS~23448+6010
	&23 47 20.07  &+60 27 21.1    &0102+584	  &$1.753\pm0.003$   &$10.0\times8.7$		&\phpm\phn4	&+5	&\phnn30\phe\\
\hline
\end{tabular}
\begin{list}{}{}
\item[$^\mathrm{a}$] rms limited by dynamic range.
\end{list}
\end{center}
\label{t3cmobs}
\end{table*}
%-------------------------------------------------------------------------------

%-------------------------------------------------------------------------------
\begin{table*}
\caption{Parameters of the observations at 1.3~cm with the VLA}
\begin{center}
\begin{tabular}{lcccccccc}
\noalign{\smallskip}
\hline\hline\noalign{\smallskip}
&\multicolumn{2}{c}{Phase Center}
&Phase
&Bootstrapped Flux
&Beam
&P.A.
&
&rms
\\
\cline{2-3}\noalign{\smallskip}
Region
&$\alpha$(J2000.0)
&$\delta$(J2000.0)
&Calibrator
&(Jy)		
&(arcsec)
&($\degr$)
&Robust
&($\mu$Jy~beam$^{-1}$)
\\
\noalign{\smallskip}
\hline\noalign{\smallskip}
IRAS~00117+6412
	 &00 14 27.73  &+64 28 46.2    &0102+584   &$2.256\pm0.015$  &$5.9\times4.5$	&$-$21		&+5	&\phnn67\phe\\
IRAS~04579+4703
	 &05 01 39.73  &+47 07 23.1    &0423+418   &$1.617\pm0.017$  &$5.4\times4.4$	&\phpm45	&+5	&\phnn85\phe\\
IRAS~18123$-$1203
	 &18 15 07.33  &$-$12 02 42.0  &1832$-$105 &$0.913\pm0.015$  &$6.9\times4.6$	&$-$20		&+5	&\phnn63\phe\\
IRAS~18171$-$1548
	 &18 20 04.62  &$-$15 46 46.6  &1832$-$105 &$0.913\pm0.015$  &$8.3\times5.0$	&$-$34		&+5	&\phn112\phe\\
IRAS~19045+0813
	 &19 06 59.77  &+08 18 42.8    &1922+155   &$0.504\pm0.002$  &$6.0\times4.2$	&\phpm41	&+4	&\phnn65\phe\\
IRAS~20293+4007
	 &20 31 07.92  &+40 17 22.8    &2025+337   &$1.946\pm0.013$  &$5.1\times4.3$	&\phpm41	&+5	&\phnn80\phe\\
IRAS~22187+5559
	&22 20 34.89  &+56 14 39.5    &2137+510   &$0.626\pm0.008$  &$7.9\times6.5$	&\phpm50	&+5	&\phnn60\phe\\
IRAS~22198+6336
	 &22 21 27.62  &+63 51 42.2    &2230+697   &$0.601\pm0.005$  &$3.6\times2.9$	&\phn$-$5	&+1	&\phnn63\phe\\
NGC\,7538-IRS9
	&23 14 01.68  &+61 27 19.9 &2230+697   &$0.601\pm0.005$  &$3.7\times3.1$	&\phpm14	&$-$5	&\phn204\phe\\
IRAS~23448+010
	 &23 47 20.07  &+60 27 21.1    &0102+584   &$2.256\pm0.015$  &$5.3\times4.3$	&\phpm40	&+5	&\phnn76\phe\\
\hline
\end{tabular}
\end{center}
\label{t1cmobs}
\end{table*}
%-------------------------------------------------------------------------------

%-------------------------------------------------------------------------------
\begin{table*}
\caption{Parameters of the observations at 7~mm with the VLA}
\begin{center}
\begin{tabular}{lcccccccc}
\noalign{\smallskip}
\hline\hline\noalign{\smallskip}
&\multicolumn{2}{c}{Phase Center}
&Phase
&Bootstrapped Flux
&Beam
&P.A.
&
&rms
\\
\cline{2-3}\noalign{\smallskip}
Region
&$\alpha$(J2000.0)
&$\delta$(J2000.0)
&Calibrator
&(Jy)		
&(arcsec)
&($\degr$)
&Robust
&($\mu$Jy~beam$^{-1}$)
\\
\noalign{\smallskip}
\hline\noalign{\smallskip}
IRAS~04579+4703
	 &05 01 39.73  &+47 07 23.1    &0502+416   &$0.424\pm0.012$  &$3.7\times3.2$	&\phpm40	&+5	&\phnn 94\phe\\
IRAS~19045+0813
	 &19 06 59.77  &+08 18 42.8    &1856+061   &$0.249\pm0.006$  &$3.9\times3.3$	&\phpm44	&+5	&\phn 100\phe\\
IRAS~22187+5559
	 &22 20 34.89  &+56 14 39.5    &2250+558   &$0.300\pm0.019$  &$3.6\times3.1$	&\phpm49	&+5	&\phn 178\phe\\
IRAS~22198+6336
	 &22 21 27.62  &+63 51 42.2    &2148+611   &$0.530\pm0.032$  &$2.2\times1.3$	&$-$43		&+0	&\phn 226\phe\\
\hline
\end{tabular}
\end{center}
\label{t7mmobs}
\end{table*}
%-------------------------------------------------------------------------------

%%%%%%%%%%%%%%%%%%%%%%%%%%%%%%%%%%%%%%%%%%%%%%%%%%%%%%%%%%%%%%%%%%%%%%%%%%%%%%%%
%
% RESULTS

\section{Results \label{sres}}

\subsection{Results at 1.2~mm \label{sresmm}}

In Table~\ref{t30m} we summarize the main results of the IRAM~30\,m
observations, in particular the peak intensity, flux density, and mass
of gas and dust for each region. In the bottom panel of
figures~\ref{fi00117_1} to \ref{fi23448_1} we show in grey scale and
black contours the 1.2~mm continuum emission. 

None of the regions observed with the IRAM~30\,m shows a point-like
morphology. On the contrary, four regions (I18123, I18171, I20293, and
I23448) have very extended emission, with no clear peak. The other
regions have a clear emission peak and show some substructure. For
I18212 no millimeter source was detected ($<620$~mJy). Thus, this
source was not observed with the VLA.

\subsection{Results at 3.6~cm, 1.3~cm, and 7~mm \label{srescm}}

In Tables~\ref{t3cm}, \ref{t1cm}, and \ref{t7mm} we list all the
sources detected in each region above the 5$\sigma$ level at 3.6~cm,
1.3~cm, and 7~mm respectively. The three tables give, for each
centimeter source identified in each region (labeled in increasing
RA), the coordinates, peak intensity, flux density, and deconvolved
size and position angle obtained by fitting a gaussian with the task
\texttt{JMFIT} in AIPS. In all three tables, we classify a VLA source
as "central" if it is located within the central $50\arcsec$ of
diameter around the IRAS source, corresponding to 0.5~pc at an average
distance of 2~kpc, this being the expected size of the cluster forming
around a high-mass source \citep{hillenbrandhartmann1998,testi1999}.
For sources with extended emission, we estimated the intensity peak
and flux density using the \texttt{TVSTAT} task in AIPS.

In the case of NGC\,7538-IRS9 at 3.6~cm, the rms was limited by
dynamic range from the strong source VLA~1 (NGC\,7538-IRS1) and the
extended emission toward the north of VLA~1 (which corresponds to an
\hii\ region detected at 21~cm, see Fig.~\ref{fi23118_1}). We
performed self-calibration in phase by using the strong source VLA~1.
Subsequently, we substracted the clean components of VLA~1 using the
task \texttt{UVSUB} in AIPS and used different \emph{uv}-ranges in
order to better clean the extended emission and reduce the rms noise.
Although the rms noise was lowered from 1.28 to 0.7~mJy~beam$^{-1}$,
this was not low enough to detect VLA~2 and VLA~3, for which
\citet{sandell2005} report a flux density of $0.76\pm0.15$~mJy.

We estimated the spectral index $\alpha$
($S_{\nu}\propto\nu^{\alpha}$) of the central sources, listed in
Table~\ref{tspindex}, using the flux densities at 3.6 and 1.3~cm. In
order to properly compare the flux at both wavelengths, we made 3.6
and 1.3~cm images using the same \emph{uv}-range (4--25~k$\lambda$) at
both wavelengths.  For the sources not detected at both wavelengths in
the 4--25~k$\lambda$ maps, we used the flux densities obtained with no
restrictions in the \emph{uv}-range, listed in Tables~\ref{t3cm} or
\ref{t1cm}. The upper-limit for non-detections was set to 4 times the
rms noise level of the map, or slightly higher for extended sources.
For details in the calculation of the spectral indices see the
footnotes of Table~\ref{tspindex}.

Additionally, we searched the NRAO VLA Sky Survey
\citep[NVSS,][]{condon1998} for emission at 21~cm, and estimated the
spectral index between 21 and 3.6~cm. We made new images at 3.6~cm
with an \emph{uv}-range of projected baselines of 1--5~k$\lambda$,
which corresponds to the \emph{uv}-range of the NVSS images. In case
of non-detection we followed the same procedure described above.

%-------------------------------------------------------------------------------
\begin{table}
\caption{Parameters of the 1.2~mm sources observed with IRAM~30\,m}
\begin{center}
{\tiny
\begin{tabular}{lcccc}
\noalign{\smallskip}
\hline\hline\noalign{\smallskip}
&rms
&$I_\nu^\mathrm{peak}$
&$S_\nu$$^\mathrm{a}$
&Mass$^\mathrm{b}$
\\
Region
&(m\jpb)
&(mJy~beam$^{-1}$)
&(mJy)
&(\mo)
\\
\noalign{\smallskip}
\hline\noalign{\smallskip}
IRAS~00117+6412		&\phn8\phs		&108		&1200  		&\phn36\\
IRAS~04579+4703		&\phn7\phs		&\phn92		&\phn390	&\phn23\\
IRAS~18123$-$1203	&15$^\mathrm{c}$	&\phn68		&\phn440	&\phn41\\
IRAS~18171$-$1548	&16$^\mathrm{c}$	&\phn74		&2200	 	&142\\
IRAS~18212$-$1320	&14$^\mathrm{c}$	&$<56$\phn	&$<620$\phn  	&\ldots\\
IRAS~19045+0813		&\phn9\phs		&100		&\phn390	&\phn10\\
IRAS~20293+4007		&\phn9\phs		&\phn55		&\phn570	&\phn64\\
IRAS~22187+5559		&			&		&		&\\
\phnn VLA~3$^\mathrm{d}$&\phn7\phs		&\phn46		&\phn180	&\phn14\\
\phnn VLA~5$^\mathrm{d}$&\phn7\phs		&\phn86		&\phn390	&\phn32\\
IRAS~23448+6010		&11\phs			&\phn63		&1600		&\phn63\\
\hline
\end{tabular}
\begin{list}{}{}
\item[$^\mathrm{a}$] Flux density inside an aperture of $80''$ of
diameter, except for IRAS 22187+5559, where the aperture was $50''$
for the source associated with VLA~5 and $30''$ for the source
associated with VLA~3.
\item[$^\mathrm{b}$] Mass of gas and dust derived assuming $\Td=30$~K,
and a dust mass opacity coefficient of 0.9~\cmg\ at 1.2~mm
\citep{ossenkopfhenning1994}.
\item[$^\mathrm{c}$] High rms noise due to the low elevation of the sources at
the time of the observations.
\item[$^\mathrm{d}$] See Section~\ref{i22187}.
\end{list}
}
\end{center}
\label{t30m}
\end{table}
%-------------------------------------------------------------------------------

For the four regions additionally observed at 7~mm (I04579, I19045,
I22187, and I22198), we detected emission in all the cases (see
Table~\ref{t7mm}), with I19045 being the faintest source. For all the
sources, the 7~mm emission was partially resolved. We estimated the
contribution of ionized gas to the 7~mm emission by using the spectral
index from 3.6 to 1.3~cm (Table~\ref{tspindex}) and extrapolating the
flux density at 1.3~cm (Table~\ref{t1cm}) to 7~mm. From the estimated
free-free flux density at 7~mm and the total observed flux density
(from Table~\ref{t7mm}), we estimated the contribution from dust at
7~mm,  which was $\sim 38$\% for the case of I04579, $\sim 44$\% for
I22198, and negligible for I19045 and I22187 (Table~\ref{tsed}). We
additionally estimated the mass of the dust component at 7~mm assuming
a dust temperature in the 50--100~K range \citep{gomez2003}, a dust
emissivity index of 2 and a dust mass opacity coefficient of 0.9~\cmg\
at 1.2~mm \citep{ossenkopfhenning1994}.  Note that we assumed a higher
temperature for the dust emission at 7~mm than for the IRAM~30\,m
emission at 1.2~mm, because the IRAM~30\,m emission must come mainly
from the (large-scale) envelope of the massive object, while the 7~mm
emission is tracing the compact dust component possibly related with a
disk, which must be hotter than the envelope. The resulting masses for
the 7~mm dust components are listed in Table~\ref{tsed}.

Figures~\ref{fi00117_1} to \ref{fi23448_1} show the maps for each
region. For most of the regions there are two main panels showing the
$K$-band of the Two Micron All Sky Survey (2MASS), and the IRAM~30\,m
1.2~mm continuum emission, together with the VLA 3.6 and 1.3~cm
continuum emission, and 7~mm continuum emission when available. For
regions with sources beyond the central $\sim\!50\arcsec$ around the
IRAS source (Figs.~\ref{fi18123_1}, \ref{fi19045_1}, and
\ref{fi20293_1}), there is an extra top panel showing the area covered
by the primary beam at 3.6~cm ($\sim\!5'$), while for
Figs.~\ref{fi22187_1}, \ref{fi22198_1}, and \ref{fi23118_1}, the
distribution of the panels is slightly different.

%-------------------------------------------------------------------------------
\begin{table*}
\caption{Parameters of the sources detected at 3.6~cm}
\begin{center}
\footnotesize
\begin{tabular}{llccccccc}
\noalign{\smallskip}
\hline\hline\noalign{\smallskip}
&VLA
&Central
&$\alpha$(J2000.0)
&$\delta$(J2000.0)
&$I_\nu^\mathrm{peak}$ $^\mathrm{b}$
&$S_\nu$ $^\mathrm{b}$
&Deconv.Size
&P.A.
\\
Region
&Source
&Source $^\mathrm{a}$
&(~\raun~)
&(~\deun~)
&(mJy~beam$^{-1}$)
&(mJy)
&($\arcsec$)   
&($\degr$)
\\
\noalign{\smallskip}
\hline\noalign{\smallskip}
IRAS~00117+6412
	       & 2--3$^\mathrm{c}$	&Y	&00 14 28.12    &+64 28 46.2    &$1.23\pm0.04$\phe  &$\phn1.61\pm0.08$\phe &$\phn8.8\times\phn4.3$	&\phn73\\
\hline\noalign{\smallskip}
IRAS~04579+4703
	       & 1	&Y	&05 01 39.92    &+47 07 21.1    &$0.15\pm0.03$\phe  &$\phn0.15\pm0.04$\phe &$\phn3.3\times\phn0.0$	&\phn30\\
\hline\noalign{\smallskip}
IRAS~18123$-$1203
	       & 1	&N	&18 15 01.36    &$-$12 01 43.2  &$0.27\pm0.05$\phe  		&$\phn0.88\pm0.23$\phe		&$19.2\times11.1$  	  &163\\
	       & 2	&Y	&18 15 07.55    &$-$12 01 43.3  &$0.22\pm0.04$\phe		&$\phn0.45\pm0.12$\phe		&$12.2\times\phn8.1$	  &\phn19\\
	       & 3	&N	&18 15 08.01    &$-$12 05 10.0  &$2.25\pm0.09$$^\mathrm{d}$	&$\phn2.62\pm0.17$$^\mathrm{d}$ &$\phn5.4\times\phn2.9$	  &\phnn2\\
	       & 4	&Y	&18 15 08.05    &$-$12 01 58.0  &$0.23\pm0.04$\phe  		&$\phn0.53\pm0.14$\phe		&$12.4\times\phn9.5$	  &107\\
	       & 5	&Y	&18 15 08.12    &$-$12 01 38.0  &$0.30\pm0.04$\phe  		&$\phn0.58\pm0.12$\phe		&$10.4\times\phn8.2$	  &\phn81\\
	       & 6	&Y	&18 15 08.75    &$-$12 01 58.0  &$0.23\pm0.04$\phe  		&$\phn0.43\pm0.12$\phe		&$10.9\times\phn6.1$	  &104\\
	       & 7	&N	&18 15 14.42    &$-$11 59 32.0  &$0.79\pm0.16$$^\mathrm{d}$	&$\phn1.85\pm0.55$$^\mathrm{d}$ &$15.2\times\phn8.0$	  &\phn52\\
\hline\noalign{\smallskip}
IRAS~18171$-$1548
	       & 1	&Y	&18 20 06.40    &$-$15 46 44.1  &$1.57\pm0.21$\phe		&$\phn3.42\pm0.63$\phe		&$12.2\times\phn9.9$	&\phn18\\
\hline\noalign{\smallskip}
IRAS~19045+0813
	       & 1	&N	&19 06 50.61    &+08 19 20.6    &$0.50\pm0.05$$^\mathrm{d}$	&$\phn0.54\pm0.09$$^\mathrm{d}$ &$\phn4.1\times\phn0.0$	&\phn64\\
	       & 2	&Y	&19 06 58.64    &+08 18 58.3    &$0.23\pm0.05$\phe		&$\phn0.38\pm0.11$\phe		&$16.2\times\phn0.0$	&\phn21\\
	       & 3	&Y	&19 06 59.34    &+08 18 56.7    &$0.41\pm0.04$\phe		&$\phn0.97\pm0.14$\phe		&$12.3\times\phn7.2$	&\phn91\\
	       & 4	&Y	&19 06 59.63    &+08 19 08.9    &$0.54\pm0.04$\phe		&$\phn1.41\pm0.15$\phe		&$15.5\times\phn6.4$	&\phn48\\
	       & 5	&Y	&19 06 59.77    &+08 19 19.1    &$0.30\pm0.04$\phe		&$\phn0.52\pm0.11$\phe		&$10.3\times\phn2.4$	&156\\
	       & 6	&Y	&19 07 00.12    &+08 19 09.9    &$0.50\pm0.04$\phe		&$\phn1.32\pm0.15$\phe		&$13.2\times\phn7.7$	&130\\
	       & 2--6$^\mathrm{c}$&Y	&\ldots	&\ldots		&$0.56\pm0.13$\phe		&$\phn3.14\pm0.19$\phe		&\ldots			&\ldots\\
	       & 7	&Y	&19 07 00.45    &+08 18 44.5    &$0.59\pm0.05$\phe		&$\phn0.60\pm0.08$\phe		&$\phn3.4\times\phn0.0$	&152\\
\hline\noalign{\smallskip}
IRAS~20293+4007
	       & 1	&Y	&20 31 07.17    &+40 17 20.9    &$0.87\pm0.06$\phe		&$\phn2.84\pm0.24$\phe		&$12.4\times\phn8.7$	&\phn22\\
	       & 2	&Y	&20 31 07.19    &+40 17 27.1    &$0.72\pm0.06$\phe		&$\phn2.08\pm0.22$\phe		&$10.6\times\phn8.7$	&179\\
	       & 1--2$^\mathrm{c}$&Y	&\ldots	&\ldots		&$0.90\pm0.22$\phe		&$\phn3.79\pm0.31$\phe		&\ldots			&\ldots\\
	       & 3	&N	&20 31 19.11    &+40 18 09.7    &$1.39\pm0.10$$^\mathrm{d}$  &$\phn1.63\pm0.20$$^\mathrm{d}$	&$\phn4.4\times\phn0.0$	&\phn72\\
\hline\noalign{\smallskip}
IRAS~22187+5559
	       & 1	&Y	&22 20 32.01    &+56 14 53.5    &$1.59\pm0.24$\phe		&$\phn1.71\pm0.24$		&\ldots			&\ldots\\
	       & 2	&Y	&22 20 33.45    &+56 15 01.5    &$2.24\pm0.37$\phe		&$\phn3.31\pm0.37$		&\ldots			&\ldots\\	       
	       & 3	&Y	&22 20 33.67    &+56 14 26.4    &$3.25\pm0.10$\phe		&$32.58\pm1.10$\phe		&$39.0\times10.7$	&128\\
	       & 4	&Y	&22 20 34.51    &+56 14 57.6    &$3.21\pm0.10$\phe		&$20.10\pm0.72$\phe		&$19.3\times13.9$	&155\\
	       & 5	&Y	&22 20 35.58    &+56 14 45.3    &$5.59\pm0.10$\phe		&$38.18\pm0.78$\phe		&$21.3\times14.0$	&170\\
	       & 1--5$^\mathrm{c}$&Y	&\ldots	&\ldots		&$5.96\pm1.06$\phe		&\phn\php$109\pm2$\phnn\php\phe &\ldots			&\ldots\\       
\hline\noalign{\smallskip}
IRAS~22198+6336
	       & 1	&N	&22 21 24.48    &+63 52 20.2    &$0.30\pm0.02$\phe		&$\phn0.30\pm0.03$\phe		&$\phn2.2\times\phn0.0$	&\phn26\\
	       & 2	&Y	&22 21 26.68    &+63 51 38.2    &$0.59\pm0.02$\phe		&$\phn0.57\pm0.03$\phe		&$\phn1.5\times\phn0.0$	&171\\
\hline\noalign{\smallskip}
NGC\,7538-IRS9 $^\mathrm{e}$
	       & 1	&N	&23 13 45.47    &+61 28 19.7    &\php$800\pm10$$^\mathrm{d}$\phn\php	&\php$1360\pm30$$^\mathrm{d}$\php\phn	&$10.2\times\phn6.0$	&\phnn4\\
\hline\noalign{\smallskip}
IRAS~23448+6010
	       & 1	&Y	&23 47 16.29    &+60 27 13.6    &$0.39\pm0.03$\phe		&$\phn0.56\pm0.06$\phe		&$13.3\times\phn0.0$	&176\\
	       & 2	&Y	&23 47 20.08    &+60 27 22.8    &$0.40\pm0.03$\phe		&$\phn1.91\pm0.14$\phe		&$23.7\times13.4$	&151\\
	       & 4	&Y	&23 47 21.97    &+60 27 48.3    &$0.21\pm0.03$\phe		&$\phn0.53\pm0.09$\phe		&$14.8\times\phn8.6$	&\phn78\\
	       & 5	&Y	&23 47 22.47    &+60 27 12.7    &$0.29\pm0.03$\phe		&$\phn0.90\pm0.10$\phe		&$15.0\times11.7$	&112\\
	       & 6	&Y	&23 47 23.21    &+60 27 33.8    &$0.26\pm0.03$\phe		&$\phn0.60\pm0.08$\phe		&$13.4\times\phn8.6$	&\phnn7\\
	       & 2--6$^\mathrm{c}$&Y	&\ldots	&\ldots		&$0.40\pm0.10$\phe		&$\phn3.19\pm0.13$\phe		&\ldots			&\ldots\\
\hline
\end{tabular}
\begin{list}{}{}
\item[$^\mathrm{a}$] Y~=~source within the central 50~$\arcsec$ around the IRAS
source; N~=~source out of the central 50~$\arcsec$ around the IRAS source.
\item[$^\mathrm{b}$] Intensity and flux density corrected for primary beam
response.
\item[$^\mathrm{c}$] Emission from the sources included in the range.
\item[$^\mathrm{d}$] Primary beam correction factor greater than 1.5.
\item[$^\mathrm{e}$] Extended emission is found toward the north of VLA~1, with
an intensity peak of $\sim\!50$~mJy~beam$^{-1}$ and a flux density of
$\sim\!1.8$~Jy. The coordinates of the intensity peak are
$\alpha$(J2000.0)~=~23$^\mathrm{h}$13$^\mathrm{m}$32$^\mathrm{s}$.63, and
$\delta$(J2000.0)~=~+61$\degr$29$\arcmin$53$\arcsec$.7 (see Fig.~\ref{fi23118_1}).
\end{list}
\end{center}
\label{t3cm}
\end{table*}
%-------------------------------------------------------------------------------

%-------------------------------------------------------------------------------
\begin{table*}
\caption{Parameters of the sources detected at 1.3~cm}
\begin{center}
\footnotesize
\begin{tabular}{llccccccc}
\noalign{\smallskip}
\hline\hline\noalign{\smallskip}
&VLA
&Central
&$\alpha$(J2000.0)
&$\delta$(J2000.0)
&$I_\nu^\mathrm{peak}$ $^\mathrm{b}$
&$S_\nu$ $^\mathrm{b}$
&Deconv.Size
&P.A.
\\
Region
&Source
&Source $^\mathrm{a}$
&(~\raun~)
&(~\deun~)
&(mJy~beam$^{-1}$)
&(mJy)
&($\arcsec$)   
&($\degr$)
\\
\noalign{\smallskip}
\hline\noalign{\smallskip}
IRAS~00117+6412
	       & 1	&N    &00 14 22.26	&+64 28 28.7    &$0.47\pm0.08$	&$\phn0.39\pm0.14$	&unresolved		&$-$\\
	       & 2	&Y    &00 14 28.32	&+64 28 41.2    &$0.47\pm0.06$	&$\phn0.70\pm0.15$	&$\phn6.0\times\phn0.0$	&\phn57\\
	       & 3	&Y    &00 14 28.43	&+64 28 47.7    &$0.57\pm0.06$	&$\phn1.24\pm0.19$	&$\phn5.9\times\phn5.2$	&119\\
	       & 2--3$^\mathrm{c}$&Y    &\ldots	&\ldots		&$0.59\pm0.15$	&$\phn1.72\pm0.20$	&\ldots			&\ldots\\
\hline\noalign{\smallskip}
IRAS~04579+4703
	       & 1    &Y    &05 01 40.05    &+47 07 21.5    &$0.54\pm0.09$  &$\phn0.44\pm0.13$  &$\phn2.2\times\phn0.0$		&\phn31\\
\hline\noalign{\smallskip}
IRAS~18171$-$1548
	       & 1	&Y    &18 20 06.46    &$-$15 46 39.9  &$0.66\pm0.12$  &$\phn2.77\pm0.62$  &$14.3\times\phn9.1$		&127\\
\hline\noalign{\smallskip}
IRAS~19045+0813
	       & 7	&Y    &19 07 00.59    &+08 18 44.1    &$0.75\pm0.05$  &$\phn1.10\pm0.12$  &$\phn5.3\times\phn0.0$	&143\\
\hline\noalign{\smallskip}
IRAS~20293+4007
 	       & 2	&Y    &20 31 07.15    &+40 17 28.1    &$0.55\pm0.08$  &$\phn1.08\pm0.22$  &$\phn6.1\times\phn3.1$	&\phn76\\
\hline\noalign{\smallskip}
IRAS~22187+5559
	       & 1	&Y    &22 20 31.87    &+56 14 55.3    &$0.70\pm0.05$  &$\phn1.32\pm0.14$  &$\phn8.4\times\phn5.0$	&\phn12\\
	       & 2	&Y    &22 20 33.38    &+56 15 06.8    &$1.13\pm0.05$  &$\phn3.79\pm0.22$  &$13.2\times\phn8.8$		&179\\	       
	       & 3	&Y    &22 20 33.72    &+56 14 29.1    &$1.99\pm0.05$  &$17.33\pm0.51$	  &$38.1\times\phn8.3$		&127\\
	       & 4	&Y    &22 20 34.59    &+56 15 00.4    &$1.95\pm0.05$  &$\phn6.05\pm0.21$  &$11.2\times\phn9.3$		&\phn99\\
	       & 5	&Y    &22 20 35.64    &+56 14 46.4    &$4.19\pm0.05$  &$19.31\pm0.28$	  &$16.0\times11.2$		&170\\
	       & 1--5$^\mathrm{c}$&Y    &\ldots	&\ldots	      &$4.64\pm0.78$  &$39.32\pm1.34$	  &\ldots			&\ldots\\       
\hline\noalign{\smallskip}
IRAS~22198+6336
	       & 2	&Y    &22 21 26.87    &+63 51 37.2    &$0.49\pm0.04$  &$\phn1.11\pm0.13$  &$\phn6.2\times\phn1.4$	&143\\
\hline\noalign{\smallskip}
NGC\,7538-IRS9
	       & 2	&Y    &23 14 01.10    &+61 27 18.8    &$1.08\pm0.20$  &$\phn1.43\pm0.43$  &$\phn4.6\times\phn0.0$	&\phn13\\
	       & 3	&Y    &23 14 01.79    &+61 27 20.0    &$2.12\pm0.20$  &$\phn3.83\pm0.52$  &$\phn4.4\times\phn1.8$	&\phn14\\
\hline\noalign{\smallskip}
IRAS~23448+6010
	       & 3	&Y    &23 47 20.61    &+60 27 16.0    &$0.63\pm0.08$  &$\phn0.80\pm0.15$  &$\phn4.2\times\phn0.0$	&154\\
\hline
\end{tabular}
\begin{list}{}{}
\item[$^\mathrm{a}$] Y~=~source within the central 50~$\arcsec$ around
the IRAS source; N~=~source out of the central 50~$\arcsec$ around the
IRAS source.
\item[$^\mathrm{b}$] Intensity and flux density corrected for primary beam
response.
\item[$^\mathrm{c}$] Emission from the sources included in the range.
\end{list}
\end{center}
\label{t1cm}
\end{table*}
%-------------------------------------------------------------------------------

%-------------------------------------------------------------------------------
\begin{table*}
\caption{Parameters of the sources detected at 7~mm}
\begin{center}
\footnotesize
\begin{tabular}{llccccccc}
\noalign{\smallskip}
\hline\hline\noalign{\smallskip}
&VLA
&Central
&$\alpha$(J2000.0)
&$\delta$(J2000.0)
&$I_\nu^\mathrm{peak}$ $^\mathrm{b}$
&$S_\nu$ $^\mathrm{b}$
&Deconv.Size
&P.A.
\\
Region
&Source
&Source $^\mathrm{a}$
&(~\raun~)
&(~\deun~)
&(mJy~beam$^{-1}$)
&(mJy)
&($\arcsec$)   
&($\degr$)
\\
\noalign{\smallskip}
\hline\noalign{\smallskip}
IRAS~04579+4703
	       & 1    &Y    &05 01 40.01    &+47 07 20.5    &$0.67\pm0.08$  &$\phn1.45\pm0.26$  &$\phn5.0\times\phn2.5$	&\phn51\\
\hline\noalign{\smallskip}
IRAS~19045+0813
	       & 7	&Y    &19 07 00.51    &+08 18 44.3    &$0.58\pm0.11$  &$\phn0.70\pm0.20$  &$\phn3.1\times\phn0.0$	&\phn31\\
\hline\noalign{\smallskip}
IRAS~22187+5559 $^\mathrm{c}$
	       & 3	&Y   &22 20 33.55    &+56 14 31.7    &$1.09\pm0.28$  &$\phn7.87\pm0.48$  &\ldots			&\ldots\\	% PBCOR map
	       & 5a	&Y   &22 20 35.27    &+56 14 36.2    &$1.10\pm0.32$  &$\phn2.81\pm0.51$	 &$\phn8.1\times\phn2.7$	&\phn30\\	% PBCOR map
	       & 5b	&Y   &22 20 35.56    &+56 14 45.4    &$1.42\pm0.33$  &$\phn3.02\pm0.63$	 &$\phn6.6\times\phn2.8$	&110\\		% PBCOR map
	       & 5c	&Y   &22 20 36.10    &+56 14 41.7    &$0.99\pm0.16$  &$\phn1.70\pm0.66$	 &$\phn6.7\times\phn4.8$	&\phnn5\\	% PBCOR map
	       & 5d	&Y   &22 20 36.27    &+56 14 38.1    &$1.07\pm0.25$  &$\phn1.71\pm0.59$	 &$\phn4.6\times\phn3.2$	&140\\		% PBCOR map
	       & 5a--5d$^\mathrm{d}$	&Y   &\ldots	     &\ldots	     &$1.42\pm0.31$  &$12.05\pm0.48$	 &\ldots	&\ldots\\	% PBCOR map
\hline\noalign{\smallskip}
IRAS~22198+6336
	       & 2	&Y    &22 21 26.76    &+63 51 37.9    &$2.10\pm0.26$  &$\phn2.76\pm0.53$  &$\phn1.5\times\phn0.3$	&110\\		% pbman 1.02
\hline
\end{tabular}
\begin{list}{}{}
\item[$^\mathrm{a}$] Y~=~source within the central 50~$\arcsec$ around
the IRAS source.
\item[$^\mathrm{b}$] Intensity and flux density corrected for primary beam
response.
\item[$^\mathrm{c}$] VLA~5a to VLA~5d are the subclumps found toward the position of VLA~5.
\item[$^\mathrm{d}$] Emission from the sources included in the range.
\end{list}
\end{center}
\label{t7mm}
\end{table*}
%-------------------------------------------------------------------------------

\subsection{Infrared photometry from 2MASS and IRAC-\emph{Spitzer} \label{sres2mass}}

For each region, we downloaded the sources of the 2MASS catalog within
the VLA primary beam at 3.6~cm, and we estimated for each 2MASS source
the infrared excess from the $(J-H)$ vs $(H-K)$ diagram, measured as
the difference between the $(H-K)$ color and the $(H-K)$ color
corresponding to a reddened main-sequence star (following the
reddening law of \citealt{riekelebofsky1985}).  We assigned a color
and a different star symbol to each infrared excess interval:
three--points blue stars for infrared excesses $<-0.4$; four--points
green stars for the interval ($-0.4$--0), corresponding to
main-sequence stars, giants, and Class III and II sources with small
infrared excess; five--points yellow stars for (0--0.4), corresponding
to Class II sources; six--points orange stars for (0.4--1),
corresponding to Class I sources, and ten--points red stars for
infrared excesses $>1$ \citep{matsuyanagi2006}. The 2MASS sources are
marked as stars with the corresponding color in Fig.~\ref{fi00117_1}
to Fig.~\ref{fi23448_1}. In our sample, I04579 and NGC\,7538-IRS9 are
the two regions where the 2MASS source associated with the IRAS source
shows the largest infrared excess. Large infrared excesses are
interpreted as reflecting an early evolutionary stage not only for
low-mass YSOs \citep[\eg][]{strom1993, kenyon1995}, but also for
intermediate/high-mass YSOs \citep[\eg][]{massi1999, nurnberger2003}.

In addition, I18171 and I19045 were observed with IRAC-\emph{Spitzer},
whose images are shown in \citet{kumargrave2007}.  For these two
regions, we studied the IRAC sources within the central $120''$ around
the \uchii\ region, and plotted the sources in $K$-[3.6] vs
[3.6]-[4.5] and [3.6]-[4.5] vs [4.5]-[5.8] diagrams, allowing us to
distinguish between Class II/III and Class 0/I sources
\citep{allen2004, hartmann2005, getman2007}.  For simplicity, in
Figs.~\ref{fi18171_1} to \ref{fi19045_1} we plotted only the IRAC
sources classified as Class 0/I sources and with no 2MASS counterpart.
We contrasted the classification of YSOs from IRAC colors with the
classification from $JHK$ infrared excess (described above) and found
that in general IRAC and 2MASS colors yield a similar classification.
Thus, for the regions not observed with IRAC we consider the
classification of YSOs from $JHK$ infrared excess as a first valid
approach to the evolutionary stage of each source.

\subsection{Spectral Energy Distributions \label{sressed}}

We searched the literature for data from 2MASS,  IRAC-\emph{Spitzer},
MSX, and SCUBA \citep{difrancesco2008} for each region. We found an
infrared counterpart of the IRAS source for all the cases except for
I22198. With these gathered data, and the data presented in this work,
we built the SED for each IRAS source, and fitted a modified blackbody
law for the far-infrared to millimeter data, taking into account the
contribution of a free-free optically thin law for the centimeter
data. Note that we did not attempt to fit the infrared excess
coming from components with higher temperatures. We adopted a dust
mass opacity coefficient at 1.2~mm of 0.9~\cmg\
\citep{ossenkopfhenning1994}, a gas-to-dust mass ratio of 100, and
assumed that the opacity follows a power law. The modified blackbody
law allowed us to estimate a range of masses, dust temperatures, and
dust emissivity indices associated with the envelope of each
intermediate/high-mass YSO (see Table~\ref{tsed} and
Fig.~\ref{fi_sed}). The median values were $\sim\!25$~\mo,
$\sim\!29$~K, and 1.9, respectively. Dust temperatures and dust
emissivity indices obtained in this work are similar to those obtained
by \citet{mookerjea2007} toward a sample of high-mass protostellar
objects located in the outer Galaxy. For I20293 and I23448 we did not
build the SED because the millimeter peak and the \uchii\ region
(associated with the IRAS source for I23448) are separated $\sim 25''$
and $40''$, respectively, and could be tracing different objects.

Regarding the contribution of the free-free emission, we adopted the
spectral index listed in Table~\ref{tspindex} (for NGC\,7538-IRS9 the
spectral index used in the fit corresponds to $-0.1$). For I22198 we
downloaded 21~cm and 6~cm VLA continuum data from the archive (project
AO145) and reduced it. At the position of VLA~2 we found a compact
source at 21~cm and 6~cm, with a flux density of $0.23\pm0.05$~mJy and
$0.38\pm0.03$~mJy, respectively. These values are in agreement with
those obtained by us at 3.6~cm, 1.3~cm and 7~mm (see the SED in
Fig.~\ref{fi_sed}).

%-------------------------------------------------------------------------------
\begin{table*}[t]
\caption{Spectral indices of the central sources}
\begin{center}
\footnotesize
\begin{tabular}{llccc}
\hline\hline\noalign{\smallskip}
&VLA
&Spectral Index$^\mathrm{a}$
&Spectral Index$^\mathrm{b}$
&
\\
Region
& Source
&(21--3.6~cm)
&(3.6--1.3~cm)
&Nature$^\mathrm{c}$
\\
\noalign{\smallskip}
\hline\noalign{\smallskip}
IRAS~00117+6412
& 2--3$^\mathrm{d}$	&$>+0.13$\phe		&$+0.18\pm0.20$\phe			&\uchii
\\

\hline\noalign{\smallskip}
IRAS~04579+4703
& 1			&$>-0.2\phn$\phe	&$+1.1\phn\pm0.4$$^\mathrm{e}$\phn	&\uchii\ / radiojet		
\\

\hline\noalign{\smallskip}
IRAS~18123$-$1203
& 2			&\ldots			&$<+0.9$$^\mathrm{e}$			&
\\

& 4			&\ldots			&$<+0.8$$^\mathrm{e}$			&
\\

& 5			&\ldots			&$<-0.4$$^\mathrm{e}$			&
\\

& 6			&\ldots			&$<-0.1$$^\mathrm{e}$			&
\\

& 2--6$^\mathrm{d}$	&$>+0.17$\phe		&\ldots					&
\\

\hline\noalign{\smallskip}
IRAS~18171$-$1548
& 1			&$>-0.3\phn$\phe	&$-0.3\phn\pm0.2$\phn\phe		&compact \hii
\\

\hline\noalign{\smallskip}
IRAS~19045+0813
& 2			&\ldots			&$<+1.1$$^\mathrm{e}$			&
\\

& 3			&\ldots			&$<+0.0$$^\mathrm{e}$			&
\\

& 4			&\ldots			&$<-0.3$$^\mathrm{e}$			&
\\

& 5			&\ldots			&$<+0.3$$^\mathrm{e}$			&
\\

& 6			&\ldots			&$<-0.2$$^\mathrm{e}$			&
\\

& 2--6$^\mathrm{d}$	&$>-0.03$$^\mathrm{f}$	&\ldots					&
\\

& 7			&$>+0.3\phn$\phe	&$+0.4\phn\pm0.2$\phn\phe		&\uchii\ / radiojet
\\

\hline\noalign{\smallskip}
IRAS~20293+4007
& 1			&\ldots			&$<-0.8$\phe				&
\\

& 2			&\ldots			&$+0.1\phn\pm0.2$\phn\phe 		&compact \hii
\\

& 1--2$^\mathrm{d}$	&$>-0.2\phn$\phe	&\ldots					&
\\

\hline\noalign{\smallskip}
IRAS~22187+5559
& 1			&\ldots			&$-0.3\phn\pm0.2$$^\mathrm{e}$\phn	&ext. ionized?
\\

& 2			&\ldots			&$+0.14\pm0.13$$^\mathrm{e}$		&ext. ionized?
\\

& 3			&\ldots			&$+0.0\phn\pm0.3$\phn\phe		&compact \hii
\\

& 4			&\ldots			&$+0.6\phn\pm0.5$\phn\phe		&ext. ionized?
\\

& 5			&\ldots			&$-0.06\pm0.17$\phe			&compact \hii
\\

& 1--5$^\mathrm{d}$	&$+0.04\pm0.02$\phe	&$+0.13\pm0.09$\phe			&
\\

\hline\noalign{\smallskip}
IRAS~22198+6336
& 2			&$+0.5\phn\pm0.2$$^\mathrm{g}$\phn	&$+0.5\phn\pm0.3\phn$\phe		&\uchii\ / radiojet
\\

\hline\noalign{\smallskip}
NGC\,7538-IRS9
& 2			&\ldots			&$>-1.2$$^\mathrm{e}$			&\uchii\ / radiojet
\\

& 3			&\ldots			&$>+0.1$$^\mathrm{e}$			&\uchii\ / radiojet
\\

& 2--3$^\mathrm{d}$	&$>-0.9$$^\mathrm{f}$	&\ldots					&
\\ %Sandell et al. 2005 (0.76 +/- 0.15) mJy

\hline\noalign{\smallskip}
IRAS~23448+6010
& 1			&\ldots			&$<-0.2$\phe				&\uchii
\\

& 2			&\ldots			&$<-0.2$\phe				&\uchii 
\\

& 3			&\ldots			&$>-0.1$\phe				&\uchii 
\\

& 4			&\ldots			&$<+1.2$$^\mathrm{e}$			&
\\

& 5			&\ldots			&$<+0.8$$^\mathrm{e}$			&
\\

& 6			&\ldots			&$<+0.9$$^\mathrm{e}$			&
\\

& 1--6$^\mathrm{d}$	&$>+0.2\phn$\phe	&\ldots					&
\\

\hline
\end{tabular}
\begin{list}{}{}
\item[$^\mathrm{a}$] Spectral index estimated from the flux density values at
21~cm from the NVSS \citep{condon1998} and at 3.6~cm (present work) obtained
with the \emph{uv}-range common to both wavelengths, 1--5~k$\lambda$. In case
of  non-detection at one of the frequencies, an upper limit of 4 times the rms
noise of the map was used.
\item[$^\mathrm{b}$] Spectral index estimated from the flux density values at
3.6~cm and 1.3~cm (present work) obtained with the \emph{uv}-range common to
both wavelengths, 4--25~k$\lambda$. In case of non-detection at one of the
frequencies, an upper limit of 4 times the rms noise of the map was used.
\item[$^\mathrm{c}$] See main text.
\item[$^\mathrm{d}$] Emission from the sources included in the range.
\item[$^\mathrm{e}$] Non-detection at 3.6~cm and 1.3~cm in the same
\emph{uv}-range maps. Flux density at 3.6~cm or 1.3~cm from Tables~\ref{t3cm} or
\ref{t1cm}. Upper limits at 3.6~cm or 1.3~cm (for the case of extended sources)
as $4A^{0.80}$ times the rms of the map (from Tables~\ref{t3cmobs} or
\ref{t1cmobs}), with $A=1+\Omega_{\mathrm{s}}/\Omega_{\mathrm{b}}$, where
$\Omega_{\mathrm{s}}$ is the solid angle of the source at the frequency where it
is detected, and $\Omega_{\mathrm{b}}$ is the beam solid angle
\citep[see][]{beltran2001}.
\item[$^\mathrm{f}$] Non-detection at 21~cm and 3.6~cm in the same
\emph{uv}-range maps. Flux density at 3.6~cm for I19045 from Table~\ref{t3cm},
and flux density at 3.6~cm of $0.76\pm0.15$~mJy for NGC\,7538-9 from
\citet{sandell2005}.
\item[$^\mathrm{g}$] Flux density at 21~cm of $0.23\pm0.05$~mJy from the VLA
archive data. See Section~\ref{sressed}.
\end{list}
\end{center}
\label{tspindex}
\end{table*}
%-------------------------------------------------------------------------------

%-------------------------------------------------------------------------------
\begin{table*}[t]
\caption{Physical parameters of the \hii\ regions}
\begin{center}
\footnotesize
\begin{tabular}{llcccccccc}
%\noalign{\smallskip}
\hline\hline\noalign{\smallskip}
&VLA
&$\nu$
&Diameter
&$T_\mathrm{B}$
&$f\,EM$$^\mathrm{a}$
&$f\,n_\mathrm{e}$$^\mathrm{a}$
&$M_\mathrm{i}$$^\mathrm{b}$
&$\dot{N}_\mathrm{i}$
&Spectral
\\
Region
&Source
&(GHz)
&(pc)
&(K)
&($10^{3}$cm$^{-6}$~pc)
&($10^{3}$cm$^{-3}$)
&(\mo)
&(s$^{-1}$)
&Type~$^\mathrm{c}$
\\
\noalign{\smallskip}
\hline\noalign{\smallskip}
IRAS~00117+6412		& 2--3		& \phn8.5
	&\phb\ 0.054	&\phb\ 0.82	&\phnm22.3\php		&\phb\ 0.64	&\phb\ $1.3\times10^{-3}$	&$3.8\times10^{44}$	&B2\\

\hline\noalign{\smallskip}
IRAS~04579+4703		& 1		& \phn8.5
	&$<0.040$	&$>0.27$	&\phnn$>7.2$		&$>0.42$	&$<1.9\times10^{-3}$		&$6.8\times10^{43}$	&B2--B3\\

\hline\noalign{\smallskip}
IRAS~18171$-$1845	& 1		& \phn8.5
	&\phb\ 0.139	&\phb\ 0.56	&\phnm14.8\php		&\phb\ 0.33	&\phb\ $1.8\times10^{-2}$	&$1.7\times10^{45}$	&B1\\

\hline\noalign{\smallskip}
IRAS~19045+0813		& 7		& \phn8.5
	&\phb\ 0.026	&\phb\ 1.02	&\phnm27.1\php		&\phb\ 1.01	&\phb\ $9.8\times10^{-4}$	&$1.1\times10^{44}$	&B2--B3\\

\hline\noalign{\smallskip}
IRAS~20293+4007		& 2		& \phn8.5
	&\phb\ 0.158	&\phb\ 0.43	&\phnm11.8\php		&\phb\ 0.27	&\phb\ $1.9\times10^{-2}$	&$1.7\times10^{45}$	&B1\\

\hline\noalign{\smallskip}
IRAS~22187+5559		
			& 3		& \phn8.5
	&\phb\ 0.287	&\phb\ 1.53	&\phnm40.8\php  	&\phb\ 0.38	&\phb\ $1.2\times10^{-1}$	&$2.0\times10^{46}$	&B0.5\\

			& 5		& \phn8.5
	&\phb\ 0.243	&\phb\ 2.51	&\phnm66.9\php  	&\phb\ 0.53	&\phb\ $1.1\times10^{-1}$	&$2.3\times10^{46}$	&B0.5\\

\hline\noalign{\smallskip}
IRAS~22198+6336		& 2		& \phn8.5
	&$<0.010$	&$>4.97$	&$>132.0$		&$>3.74$	&$<6.4\times10^{-4}$		&$7.0\times10^{43}$	&B3\\

\hline\noalign{\smallskip}
NGC\,7538-IRS9		& 2		& 22.5
	&$<0.062$	&$>0.18$	&\phn$>39.1$		&$>0.79$	&$<1.9\times10^{-3}$		&$9.0\times10^{44}$	&B1--B2\\

			& 3		& 22.5
	&\phb\ 0.038	&\phb\ 1.27	&\phm279.0\php		&\phb\ 2.70	&\phb\ $4.0\times10^{-3}$	&$2.4\times10^{45}$	&B1\\

\hline\noalign{\smallskip}
IRAS~23448+6010		& 1		& \phn8.5
	&$<0.129$	&$>0.06$	&\phnn$>1.7$		&$>0.11$	&$<3.1\times10^{-3}$		&$1.6\times10^{44}$	&B2--B3\\

			& 2		& \phn8.5
	&\phb\ 0.173	&\phb\ 0.12	&\phn\phnm3.1\php	&\phb\ 0.14	&\phb\ $9.0\times10^{-3}$	&$5.5\times10^{44}$	&B2\\

			& 3		& 22.5
	&$<0.041$	&$>0.12$	&\phn$>25.9$		&$>0.80$	&$<7.0\times10^{-3}$		&$2.6\times10^{44}$	&B2\\

\hline
\end{tabular}
\begin{list}{}{}
\item[$^\mathrm{a}$] \emph{f} corresponds to the filling factor.
\item[$^\mathrm{b}$] Mass of ionized gas estimated from the beam averaged electron density and the observed size of the source.
\item[$^\mathrm{c}$] Estimated from \citet{panagia1973}.
\end{list}
\end{center}
\label{tphyspar}
\end{table*}
%-------------------------------------------------------------------------------

\section{Discussion \label{sdisc}}

\subsection{On the nature of the cm emission associated with IRAS sources \label{scmemission}}

Most of the sources classified as central sources in Tables~\ref{t3cm}
and \ref{t1cm} have a flat or positive spectral index (see
Table~\ref{tspindex}).  All of them (with the exception of I18123) are
nearly coincident with the corresponding IRAS source, which in all
cases has a bolometric luminosity $\gtrsim10^3$~\lo,  suggesting that
the centimeter sources associated with the IRAS source are tracing
\uchii\ regions. However, the centimeter emission from the sources
associated with molecular outflows, I22198 and NGC\,7538-IRS9, could
also arise in gas ionized by shocks in the outflow. The centimeter
continuum luminosity, $S_\nu d^2$, from shock-ionized gas is
proportional to the outflow momentum rate, $\dot{P}$, with a constant
of proportionality including an efficiency factor, $\eta$, defined as
the fraction of the stellar wind that is shocked \citep{curiel1987}.
For low-mass YSOs, \citet{anglada1995} found that the centimeter
emission can be produced by shock-ionized gas with efficiencies of
10\%. In our case, the efficiencies needed to account for the observed
centimeter emission should be much higher than those observed for
low-mass YSOs, $\sim\,90$\% for I22198 and $\sim\,50$\% for
NGC\,7538-IRS9, being reasonable to assume that, at least for these
two sources, the contribution of shock-ionized gas in these sources if
any, is only a small fraction of the observed continuum centimeter
emission. Since these two sources seem to be among the youngest in our
sample (see \S~\ref{sdis}), the contribution of shock-ionized gas in
the other sources is expected to be also a small fraction of the
observed centimeter emission. Thus, the centimeter sources associated
with the IRAS source are most likely tracing compact or \uchii\
regions, and in Table~\ref{tphyspar} we list their physical
parameters. Note that the emission measure, electron density, and
opacity are beam--averaged values, and due to our angular resolution
of $\sim\!10\arcsec$ at 3.6~cm, these values are affected by an
important filling factor.

From Table~\ref{tphyspar}, we can see that the underlying star
exciting the \uchii\ region is in all cases an early-type B star. We
note that in the estimation of the physical parameters we assumed that
the emission is optically thin. This is supported by the spectral
indices close to zero reported in Table~\ref{tspindex} for almost all
the sources, except for I04579, for which the estimation of the rate
of ionizing photons is a lower limit \citep[\eg][]{keto2008}, and the
spectral type of the ionizing star is earlier than B2--B3. For almost
all the regions, the bolometric luminosity derived from the centimeter
emission (from the spectral type and following \citealt{panagia1973})
is lower than the IRAS bolometric luminosity estimated from the
infrared emission. Only for I00117 and I22187 we found the IRAS
luminosity to be a factor of $\lesssim 2$ smaller. However, in both
cases the centimeter sources split up in different components when
observed with higher angular resolution (see \S~\ref{i00117} and
\ref{i22187}), being the 3.6~cm flux and the derived luminosity (at
least for these two sources) an upper limit to the luminosity coming
from the massive YSO. Thus, the IRAS luminosity seems to be enough to
account for the centimeter emission by photoionization. We concluded
that all the centimeter sources associated with the IRAS sources of
our sample seem to trace a compact or \uchii\ region.

\subsection{Comments on individual sources \label{scom}}

\subsubsection{IRAS~00117+6412\label{i00117}}

\begin{figure}[h!]
\begin{center}
\begin{tabular}[b]{c}
\noalign{\bigskip}
	\epsfig{file=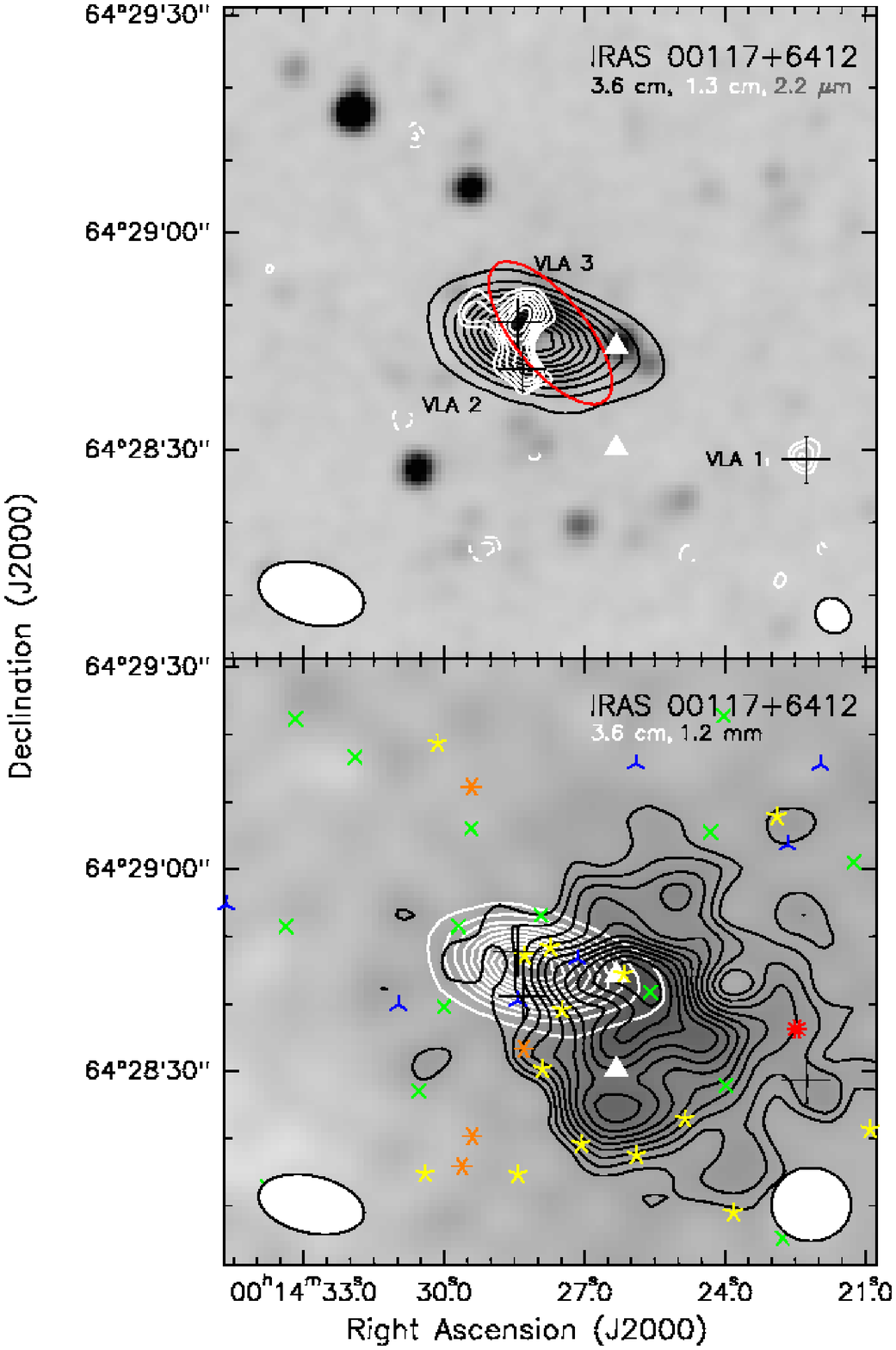, scale=0.49} \\
\noalign{\bigskip}
\end{tabular}
\caption{{\small IRAS~00117+6412.
\emph{Top}: Black contours: VLA 3.6~cm continuum emission. Levels are $-6$, $-3$
and 3 to 33 in steps of 3, times 0.038~mJy~beam$^{-1}$. White contours: VLA
1.3~cm continuum emission. Levels are $-4$, $-3$, 3, to 8 in steps of 1, times
0.067~mJy~beam$^{-1}$. Grey scale: K-band 2MASS image.
\emph{Bottom}: Grey scale and black contours: IRAM~30\,m continuum emission at
1.2~mm. Levels are 3 to 13 in steps of 1, times 8~mJy~beam$^{-1}$. White
contours: 3.6~cm emission as in top panel. Symbols in Figures~\ref{fi00117_1} to
\ref{fi23448_1} are described here. Red ellipse indicates the error ellipse of
the IRAS source. White and/or black crosses indicate the centimeter sources from
Tables~\ref{t3cm}, and \ref{t1cm}. White filled circles indicate \chtoh\ masers,
white filled triangles \water\ masers, and white filled squares OH masers. Color
stars indicate 2MASS sources with the color and symbol corresponding to
different infrared excess as described in \S~\ref{sres2mass}. Synthesized beams
at the bottom-left corners correspond to the first wavelength indicated at the
top of the panel, and beams at the bottom-right corners, when available,
correspond to the second wavelength at the top of the panel.
}}
\label{fi00117_1}
\end{center}
\end{figure}

At 3.6~cm there is one source elongated roughly in the east-west
direction, and at 1.3~cm the source splits up into two components,
with the northern one, VLA~3, nearly coincident with the peak at
3.6~cm (Fig.~\ref{fi00117_1}). The millimeter emission has the main
peak located $\sim\!15''$ to the west of VLA~3 and shows a rather
circular structure with some subcondensations. There are two \water\
masers close to the main and the secondary millimeter peaks
\citep{cesaroni1988}, suggesting that these two peaks are tracing
embedded YSOs, and there is also outflow emission centered around the
secondary millimeter peak \citep{zhang2005,kimkurtz2006}. Observations
with higher angular resolution at 3.6~cm (Busquet \et, in prep.)
reveal no emission at the position of VLA~2, and that the source
detected in this work at 3.6~cm consists of two components. One
associated with VLA~3 and with the brightest $K$-band infrared source
in the region, which has some infrared excess, and the other
coincident with the main millimeter peak (likely harboring a deeply
embedded intermediate-mass YSO). Thus, VLA~3 is most likely an \uchii\
region at the border of a dusty cloud where active star formation is
taking place.  The SED of I00117 can be fitted for temperatures
between 26 and 30~K, and for masses between 21 and 30~\mo, which are
among the median values of all the sources in this survey.

%---------------------------------------------------------------------
%---------------------------------------------------------------------

%---------------------------------------------------------------------
\begin{figure}[h!]
\begin{center}
\begin{tabular}[b]{c}
\noalign{\bigskip}
	\epsfig{file=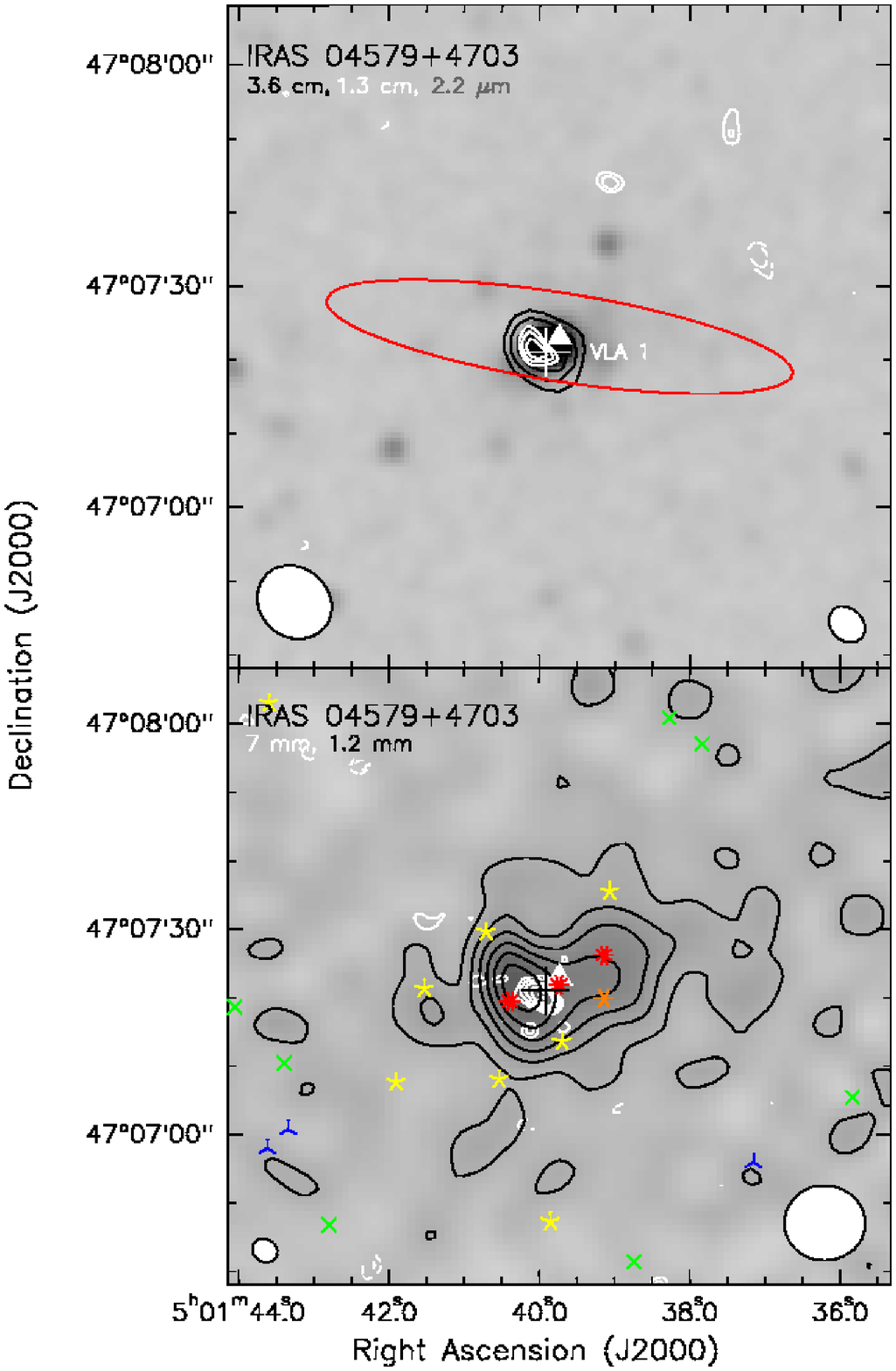, scale=0.49} \\
\noalign{\bigskip}
\end{tabular}
\caption{{\small IRAS~04579+4703.
\emph{Top}: Black contours: VLA 3.6~cm continuum emission. Levels are $-4$, $-3$
and 3 to 7 in steps of 1, times 0.022~mJy~beam$^{-1}$. White contours: VLA
1.3~cm continuum emission. Levels are $-4$, $-3$, 3, to 5 in steps of 1, times
0.085~mJy~beam$^{-1}$. Grey scale: K-band 2MASS image.
\emph{Bottom}: Grey scale and black contours: IRAM~30\,m continuum emission at
1.2~mm. Levels are 2 to 12 in steps of 2, times 7~mJy~beam$^{-1}$. White contours: VLA 7~mm
continuum emission. Levels are $-4$, $-3$, 3, to 7 in steps of 1, times
0.094~mJy~beam$^{-1}$. Color
stars indicate 2MASS sources and white filled symbols refer to masers
(see Fig.~\ref{fi00117_1} for more details).
}}
\label{fi04579_1}
\end{center}
\end{figure}
%---------------------------------------------------------------------

%---------------------------------------------------------------------
\begin{figure}[p!]
\begin{center}
\begin{tabular}[b]{c}
\noalign{\bigskip}
	\epsfig{file=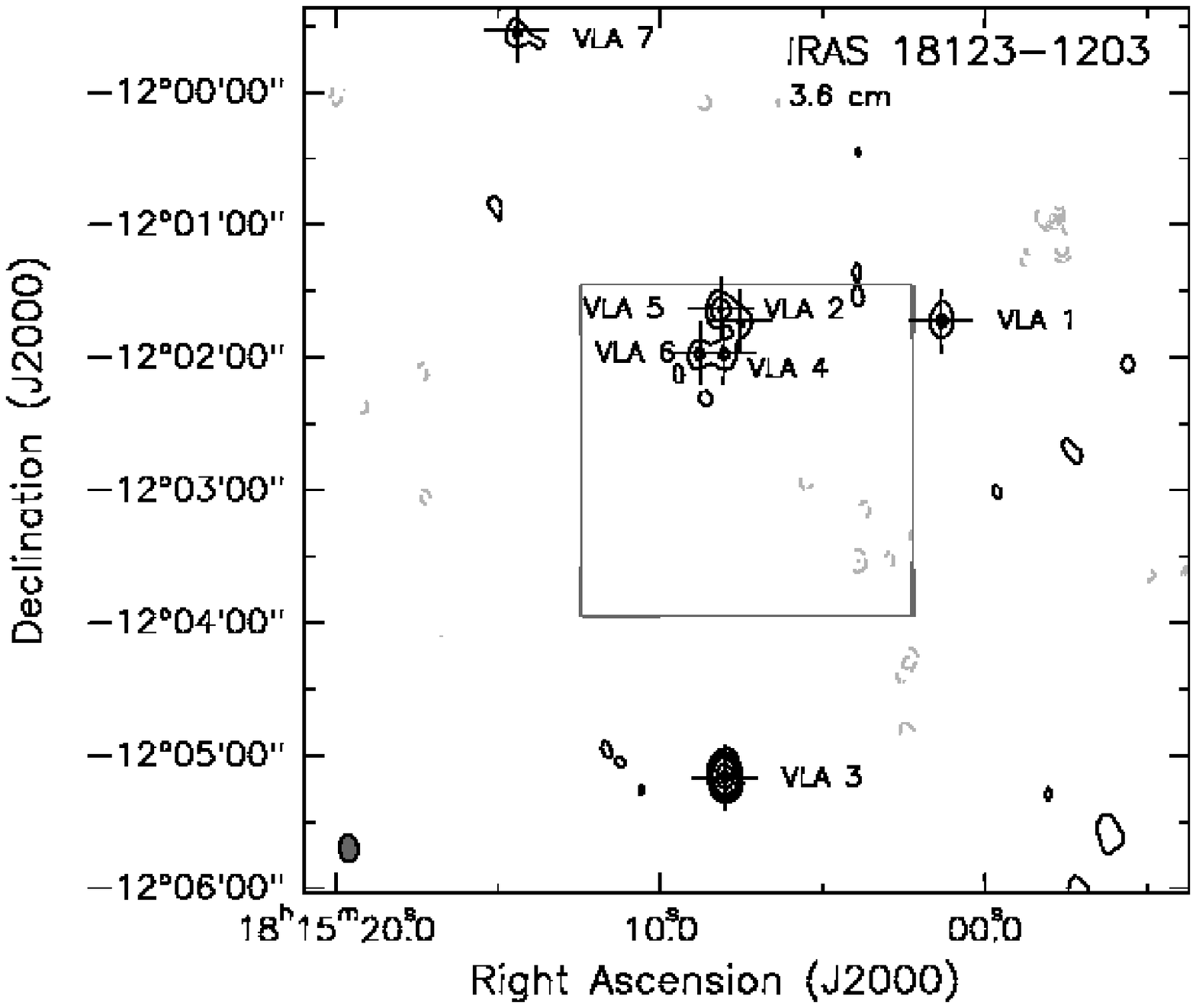, scale=0.49} \\
\noalign{\bigskip}
	\epsfig{file=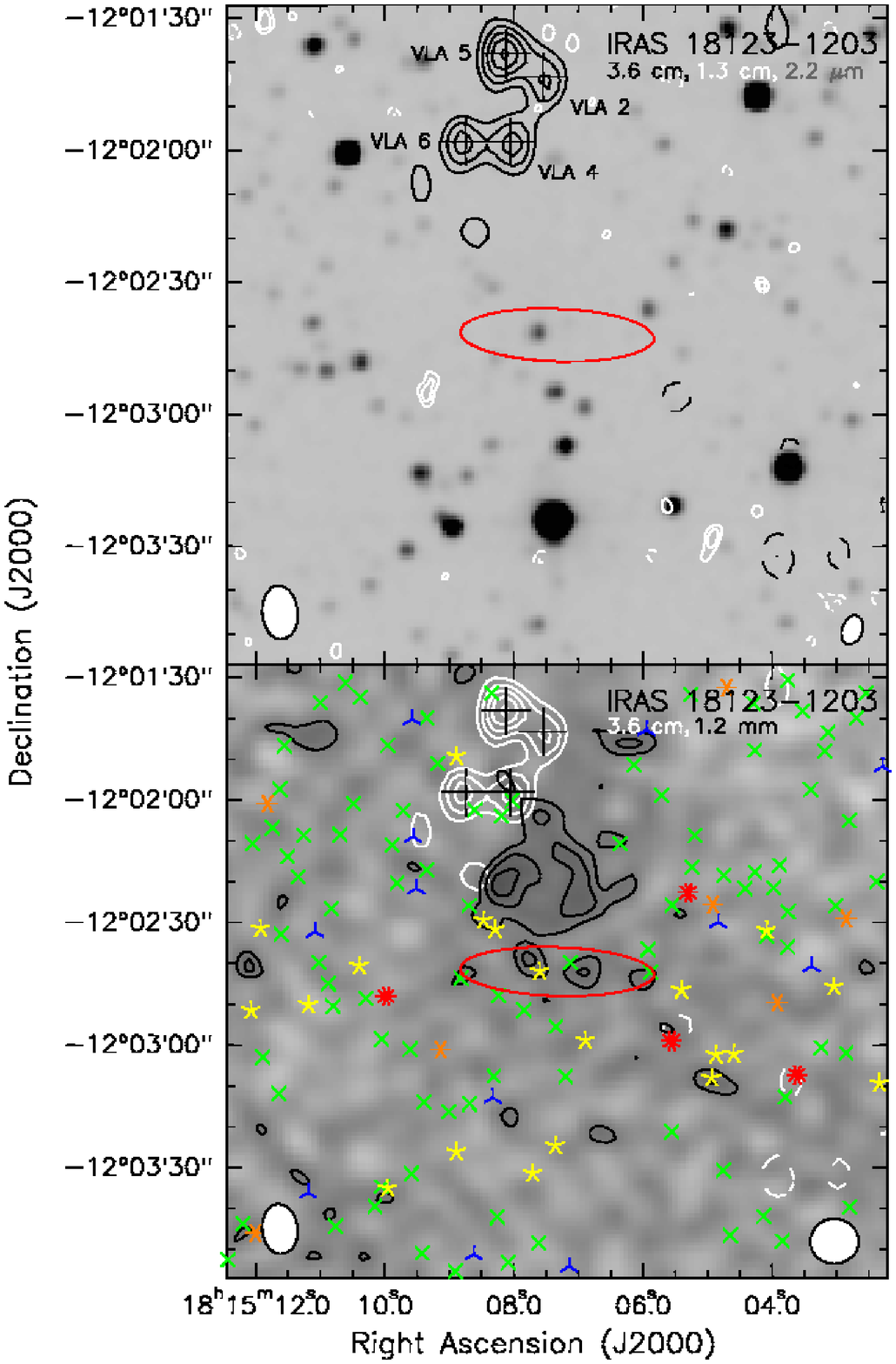, scale=0.49} \\
\noalign{\bigskip}
\end{tabular}
\caption{{\small IRAS~18123$-$1203.
\emph{Top}: VLA 3.6~cm map. The box shows the region zoomed in the
other panels.
\emph{Middle}: Black contours: VLA 3.6~cm continuum emission. Levels are $-4$,
$-3$ and 3 to 7 in steps of 1, times 0.043~mJy~beam$^{-1}$. White contours: VLA
1.3~cm continuum emission. Levels are $-4$, $-3$, 3, and 4, times
0.063~mJy~beam$^{-1}$. Grey scale: K-band 2MASS image.
\emph{Bottom}: Grey scale and black contours: IRAM~30\,m continuum emission at
1.2~mm. Levels are 2 to 5 in steps of 1, times 15~mJy~beam$^{-1}$. White contours: 3.6~cm
emission as in middle panel. Color
stars indicate 2MASS sources and white filled symbols refer to masers
(see Fig.~\ref{fi00117_1} for more details).
}}
\label{fi18123_1}
\end{center}
\end{figure}
%---------------------------------------------------------------------

%---------------------------------------------------------------------
\begin{figure}[htbp!]
\begin{center}
\begin{tabular}[b]{c}
\noalign{\bigskip}
	\epsfig{file=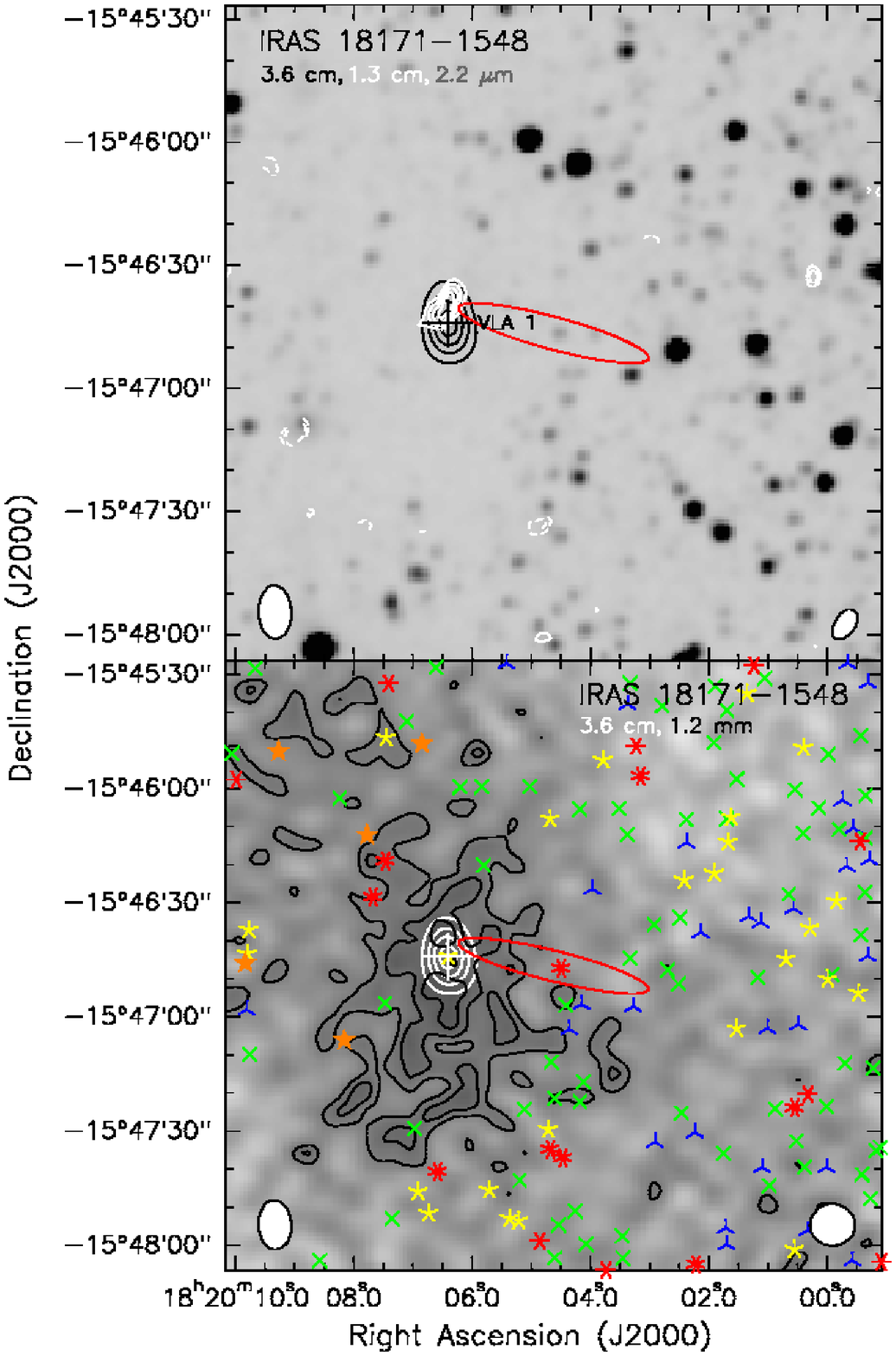, scale=0.49} \\
\noalign{\bigskip}
\end{tabular}
\caption{{\small IRAS~18171$-$1548.
\emph{Top}: Black contours: VLA 3.6~cm continuum emission. Levels are $-5$, $-3$
and 3 to 9 in steps of 2, times 0.185~mJy~beam$^{-1}$. White contours: VLA
1.3~cm continuum emission. Levels are $-4$, $-3$, 3, to 6 in steps of 1, times
0.112~mJy~beam$^{-1}$. Grey scale: K-band 2MASS image.
\emph{Bottom}: Grey scale and black contours: IRAM~30\,m continuum emission at
1.2~mm. Levels are 2 to 5 in steps of 1, times 16~mJy~beam$^{-1}$. White
contours: 3.6~cm emission as in top panel. Orange filled stars indicate
IRAC-\emph{Spitzer} sources classified as Class 0/I sources (see
\S~\ref{sres2mass}), and with no 2MASS counterpart. Color
stars indicate 2MASS sources and white filled symbols refer to masers
(see Fig.~\ref{fi00117_1} for more details).
}}
\label{fi18171_1}
\end{center}
\end{figure}
%---------------------------------------------------------------------

\subsubsection{IRAS~04579+4703\label{i04579}}

The 3.6~cm source in this region is elongated in the
southwest-northeast direction and has a counterpart at 1.3~cm,
elongated also in the same direction.  The spectral index, $\sim\!1$,
indicates that the emission is thermal and partially thick. Such a
spectral index has been found for thermal radiojets associated with
sources powering molecular outflows
\citep[\eg][]{beltran2001,zapata2004,gibbhoare2007}, and for
hypercompact \hii\ regions
\citep[\eg][]{carral1997,ignacechurchwell2004,kurtz2005}. There is an
infrared source with strong infrared excess associated with the
centimeter source, suggesting that the massive YSO is likely
surrounded by circumstellar material (see Fig.~\ref{fi04579_1}). In
addition, the dust continuum emission at 1.2~mm consists of a main
condensation whose peak is clearly associated with the centimeter
source (the offset is $<5''$, the positional uncertainty of the
IRAM~30\,m), as well as with \water\ maser and \nh\ emission
\citep{palla1991,molinari1996}. The spatial coincidence of the
millimeter peak with the centimeter source suggests that the
underlying massive star is in a very young evolutionary stage, not
having pushed away the surrounding material yet. This is reinforced by
the detection of a compact source at 7~mm at the position of the
centimeter source. Note that 38\% of the 7~mm flux density comes from
the dust emission from the envelope/disk associated with the massive
YSO, and for this dust component we estimated a mass of
$\sim\!13$~\mo. Note also that the 7~mm emission, similar to the 3.6
and 1.3~cm emission, is slightly elongated in the southwest-northeast
direction. Regarding the outflow emission, \citet{zhang2005} detect no
CO\,(2--1) outflow in the region using the NRAO\,12~m Telescope.
However, CO\,(1--0) emission had been previously detected using the
IRAM\,30~m with a linewidth of $6.1$~\kms\ by
\citet{wouterlootbrand1989}, with a sensitivity similar to that
achieved by \citet{zhang2005}. Further sensitive observations are
required to confirm or discard the presence of an outflow associated
with the massive YSO, and thus to properly interpret the nature of the
elongated emission at 7~mm. The values for the dust temperature and
mass derived from the SED fit are around 30~K and 23~\mo, which are
near the median values derived for the sources of this work.  Finally,
note that there are 2MASS sources showing infrared excess, surrounding
the centimeter source and falling within the millimeter condensation
(see Fig.~\ref{fi04579_1}), indicating that a cluster of low-mass YSOs
is possibly forming around the massive YSO.

\subsubsection{IRAS~18123$-$1203\label{i18123}}

We detected four sources toward I18123 at 3.6~cm within the central
$50\arcsec$, as well as a dust condensation consisting of different
peaks (see Fig.~\ref{fi18123_1}). However, neither the centimeter
emission nor the main millimeter emission are coincident with the IRAS
source, and the centimeter and millimeter emissions are not
overlapping. In addition, there are no 2MASS sources associated with
the main dust condensation or the centimeter sources (except maybe for
VLA~4). About $10''$ to the south of the main dust condensation there
is a faint and compact millimeter source at 3$\sigma$, associated with
a 2MASS source with some infrared excess, and within the IRAS error
ellipse. The faint $JHK$ magnitudes of this 2MASS source, its marginal
detection at 1.2~mm, and the non-detection of \nh\ emission toward the
region \citep{molinari1996} suggest that the YSO associated with the
IRAS source should be of low-mass. However, this is in conflict with
the assumed IRAS bolometric luminosity of 7900~\lo. Thus, either this
object is peculiar, or its distance is overestimated.

The spatial anticorrelation between the centimeter and the millimeter
emission is somewhat puzzling. We first considered the possibility
that the centimeter emission is tracing externally ionized gas, and
searched the catalog of \citet{reed2003}\footnote{Catalog available at
http://othello.alma.edu/$\sim$reed/OBfiles.doc} for nearby OB stars.
We did not find any OB star to the northeast of the centimeter
emission which could account (following \citealt{lefloch1997}) for the
total centimeter flux of VLA~2, VLA~4 to VLA~6 (1.6~mJy), and thus we
ruled out the possibility that these centimeter sources were
externally ionized globules. Another possibility is that the
centimeter emission traces extragalactic objects, suggested by the
double radio source morphology. However, there are no sources detected
at 21~cm in the NVSS (the spectral index between 21 and 3.6~cm being
typical of thermal emission), and the spectral indices between 3.6 and
1.3~cm do not discard a free-free thermal nature for the centimeter
emission (except for VLA~5). Thus, the centimeter emission to the
north of the IRAS source could be produced by a group of low-mass YSOs
driving thermal radiojets, whose extended emission may have been
resolved out at 1.3~cm.

%-------------------------------------------------------------------------------
\begin{table*}
\caption{Parameters of the Spectral Energy Distribution fits, and dust thermal
emission at 1.2 and 7~mm}
\begin{center}
\footnotesize
\begin{tabular}{llcccccccccc}
\hline\hline\noalign{\smallskip}
&
&
&\multicolumn{3}{c}{From SED Fit}
&
&\multicolumn{2}{c}{1.2~mm}
&
&\multicolumn{2}{c}{7~mm Dust Emission}
\\
\cline{4-6}\cline{8-9}\cline{11-12}
\noalign{\smallskip}
&VLA
&free-free
&
&$T$
&$M$
&
&\multicolumn{2}{c}{$M$ $^\mathrm{a}$}
&
&$S_\nu$ $^\mathrm{b}$
&$M$ $^\mathrm{c}$
\\
Region
&Source
&Spectral Index
&$\beta$
&(K)
&(\mo)
&
&\multicolumn{2}{c}{(\mo)}
&
&(mJy)
&(\mo)
\\
\noalign{\smallskip}
\hline\noalign{\smallskip}
IRAS~00117+6412		&2--3$^\mathrm{d}$
			&+0.18\phe	&1.5--1.7	&30--26		&\phn21--30\phn		&&\multicolumn{2}{c}{\phn36\phe}
			&&		&		\\

IRAS~04579+4703		&1			
			&+1.1\phn\phe	&1.8--2.0	&32--28		&\phn20--26\phn		&&\multicolumn{2}{c}{\phn23\phe}
			&&$0.5\pm0.4$	&8--19		\\

IRAS~18171$-$1548	&1
			&$-$0.3\phn\phe	&1.7--2.0	&30--26		&100--170		&&\multicolumn{2}{c}{142\phe}
			&&		&		\\

IRAS~19045+0813		&7
			&+0.4\phn\phe	&1.9--2.2	&29--26		&\phnn8--12\phn		&&\multicolumn{2}{c}{\phn10\phe}
			&&\ldots	&\ldots	\\

IRAS~22187+5559		&3
			&+0.0\phn\phe	&1.9--2.1	&34--32		&\phn10--20\phn		&&\multicolumn{2}{c}{\phn14\phe}
			&&\ldots	&\ldots	\\

	   		&5
			&$-$0.06\phe	&1.9--2.0	&32--31		&\phn30--40\phn		&&\multicolumn{2}{c}{\phn32\phe}
			&&\ldots	&\ldots \\

IRAS~22198+6336		&2
			&+0.5\phn\phe	&1.7--1.9	&29--25		&\phn20--30\phn		&&\multicolumn{2}{c}{\phn20$^\mathrm{e}$}
			&&$1.2\pm0.7$	&6--11		\\

NGC\,7538-IRS9		&2--3$^\mathrm{d}$
			&$-$0.10$^\mathrm{f}$	&1.9--2.0 &35--32	&\phn50--80\phn		&&\multicolumn{2}{c}{\phn66$^\mathrm{g}$}
			&&		&		\\
\hline
\end{tabular}
\begin{list}{}{}
\item[$^\mathrm{a}$] Dust and gas mass from IRAM~30\,m 1.2~mm observations (Table~\ref{t30m}).
\item[$^\mathrm{b}$] Flux density at 7~mm without the free-free contribution estimated from the 3.6 and 1.3~cm emissions.
\item[$^\mathrm{c}$] Dust and gas mass from the 7~mm dust emission,
assuming a dust temperature in the 50--100~K range \citep{gomez2003},
a dust emissivity index of 2 and a dust mass opacity coefficient of
0.9~\cmg\ at 1.2~mm \citep{ossenkopfhenning1994}. The higher value corresponds to a temperature of 50~K, and the lower to a temperature of 100~K.
\item[$^\mathrm{d}$] Emission from the sources included in the range.
\item[$^\mathrm{e}$] Estimated from the 850~\mum\ flux density from \citet{jenness1995}.
\item[$^\mathrm{f}$] Value adopted for the SED fit, due to the detection of the source at only one wavelength.
\item[$^\mathrm{g}$] Estimated from the 1.2~mm flux density from \citet{sandellsievers2004}.
\end{list}
\end{center}
\label{tsed}
\end{table*}
%-------------------------------------------------------------------------------

\subsubsection{IRAS~18171$-$1548\label{i18171}}

The centimeter source, detected at 3.6 and 1.3~cm, has a flat spectral
index, a size of $\sim\!0.15$~pc, and is clearly associated with an
infrared source detected in the $JHK$ bands from 2MASS and revealed by
the IRAC-\emph{Spitzer} images at 3.6--8.0~$\mu$m as a bright
nebulosity (but with no point source) emerging from a dark structure
\citep{kumargrave2007}.  Thus, the centimeter source is most likely
tracing a compact \hii\ region. Note that the compact \hii\ region
lies in a zone with a low density of infrared sources (see the 2MASS
$K$-band image in Fig.~\ref{fi18171_1}), whose central part is
coinciding with the dark structure revealed by IRAC. Note also that
the IRAS source is slightly shifted to the west of the compact \hii\
region, and has a 2MASS source with strong infrared excess associated.
The parameters derived for the compact \hii\ region suggest that the
underlying star is a B1 star, and from the fit to the SED we estimated
a mass of the envelope around 100~\mo\ for dust temperatures of
$\sim\!28$~K. The morphology revealed by the large-scale millimeter
emission shows an elongated and clumpy condensation in the north-south
direction coinciding with the zone with a small number of 2MASS
sources. Thus, the millimeter emission seems to trace the cloud of
dense gas and dust in which the massive star was formed, which is the
most massive cloud of this survey, as derived from the flux density at
1.2~mm.  It is worth noting that the main dust condensation contains
multiple peaks or subcondensations. Interestingly, the millimeter
emission shows two peaks separated $\sim\!15''$, with the compact
\hii\ region lying in between them. This is suggestive of the compact
\hii\ region pushing out the surrounding material toward the north and
south, with the cloud being in the process of disruption, and
decreasing its density. This could explain why no NH$_3$
\citep{molinari1996}, and CS\,(2--1) \citep{bronfman1996} has been
detected toward the region.

\subsubsection{IRAS~19045+0813\label{i19045}}

\begin{figure}[htbp!]
\begin{center}
\begin{tabular}[b]{c}
\noalign{\bigskip}
	\epsfig{file=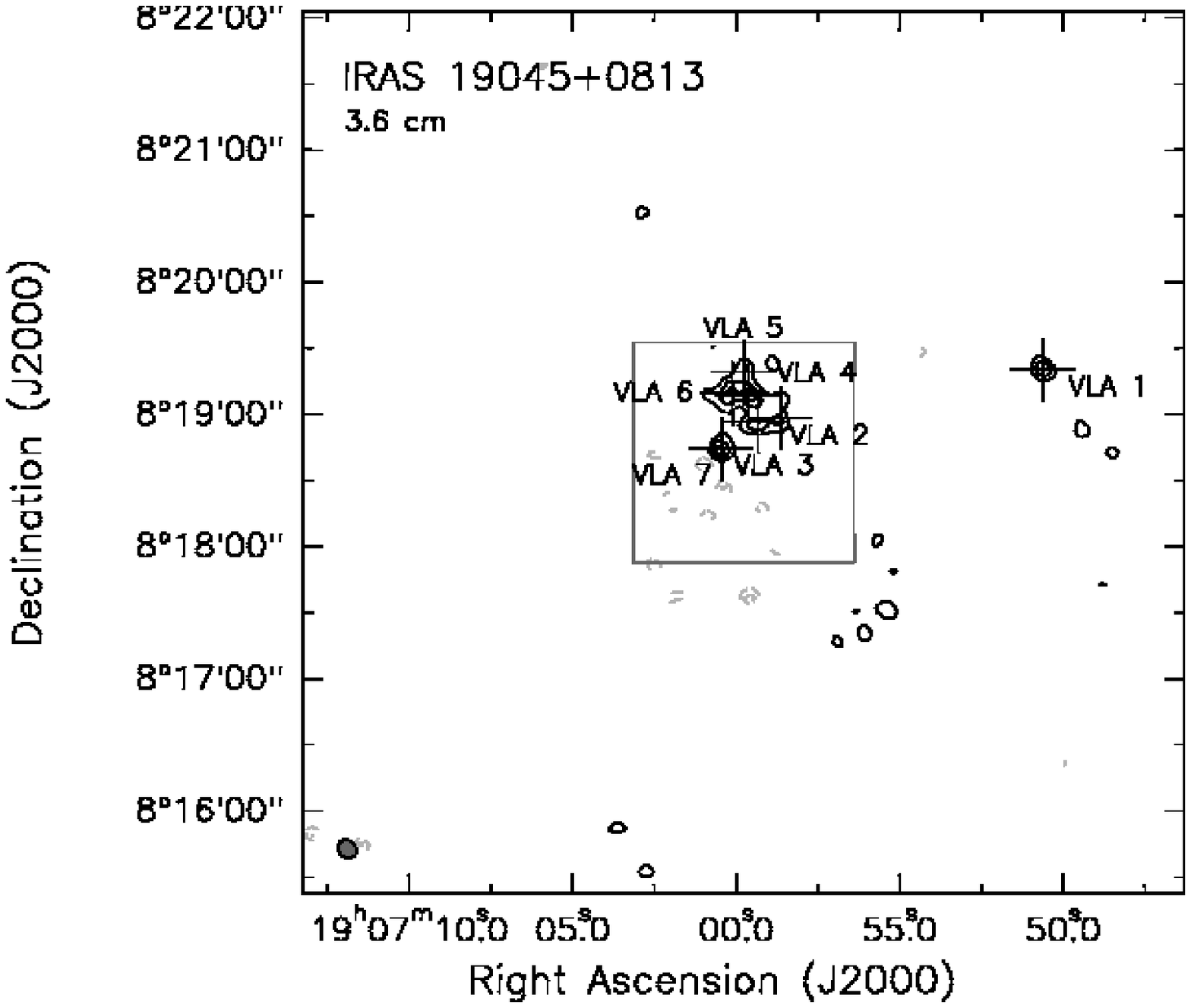, scale=0.48} \\
\noalign{\bigskip}
	\epsfig{file=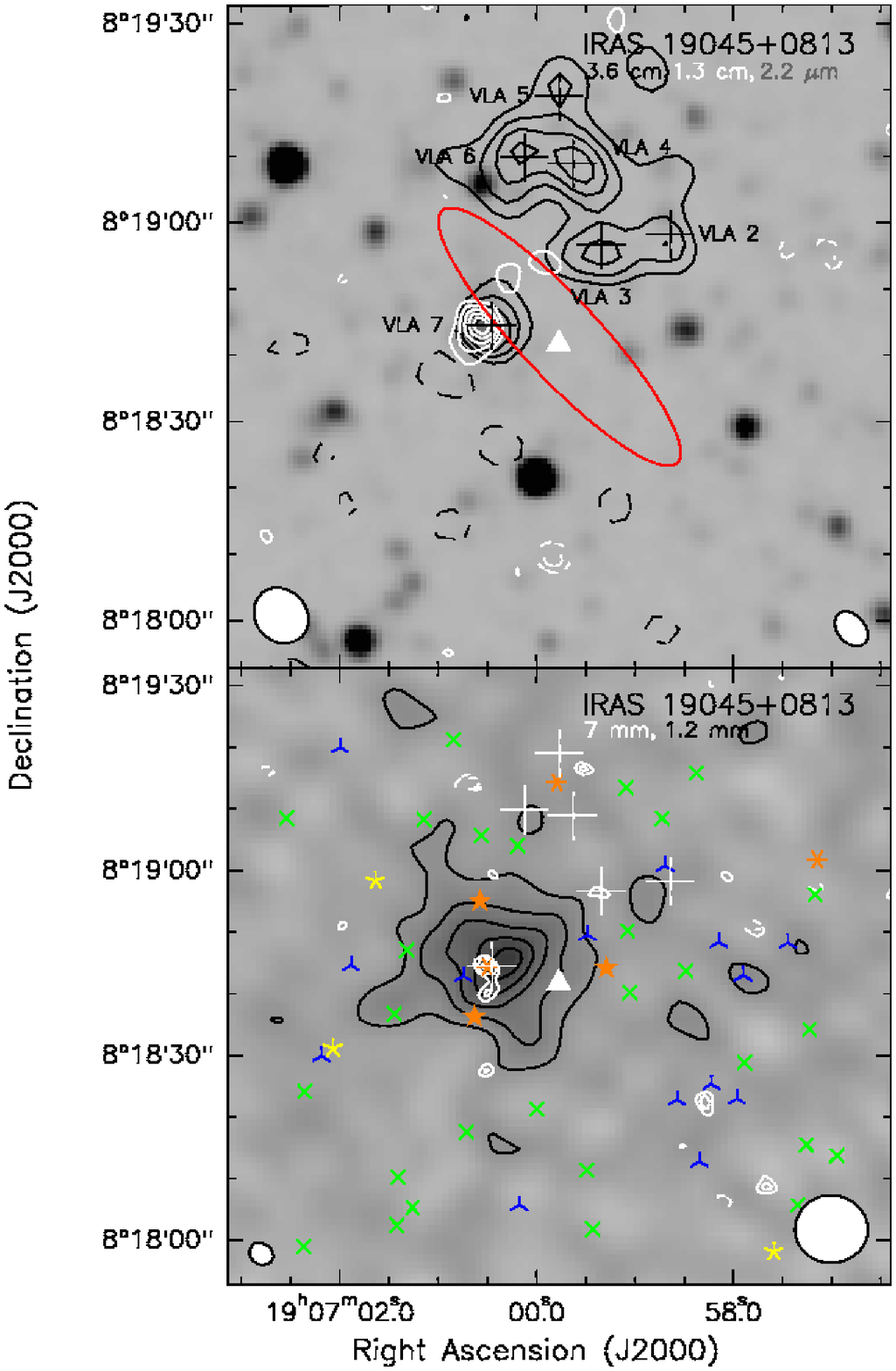, scale=0.48} \\
\noalign{\bigskip}
\end{tabular}
\caption{{\small IRAS~19045+0813.
\emph{Top}: VLA 3.6~cm map. The
box shows the region zoomed in the other panels.
\emph{Middle}: Black contours: VLA 3.6~cm continuum emission. Levels are $-5$,
$-3$ and 3 to 11 in steps of 2, times 0.054~mJy~beam$^{-1}$. White contours: VLA
1.3~cm continuum emission. Levels are $-5$, $-3$, 3 to 11 in steps of 2, times
0.065~mJy~beam$^{-1}$. Grey scale: K-band 2MASS image.
\emph{Bottom}: Grey scale and black contours: IRAM~30\,m continuum emission at
1.2~mm. Levels are 2 to 10, in steps of 2, times
9~mJy~beam$^{-1}$. White contours: VLA 7~mm
continuum emission. Levels are $-4$, $-3$ and 3 to 5 in steps of 1, times
0.100~mJy~beam$^{-1}$. Color (orange filled) stars indicate 2MASS (IRAC-Spitzer) sources and white
filled symbols refer to masers (see Figs.~\ref{fi00117_1} and \ref{fi18171_1}
for details).
}}
\label{fi19045_1}
\end{center}
\end{figure}

The centimeter emission toward I19045 presents two main components
(see Fig.~\ref{fi19045_1}), the \uchii\ region candidate, and a second
component with extended emission to the northwest of the \uchii\
region. The \uchii\ region, falling within the IRAS error ellipse, is
associated with a 2MASS source with infrared excess and with an
IRAC-\emph{Spitzer} source \citep{kumargrave2007} classified as Class
0/I source from the $K$-[3.6] vs [3.6]-[4.5] and [3.6]-[4.5] vs
[4.5]-[5.8] diagrams (\S~\ref{sres2mass}). There is CS\,(2--1)
\citep{bronfman1996} and \water\ maser emission \citep{palla1991}
detected in the region. The 1.2~mm continuum emission is associated
with the \uchii\ region, with three IRAC sources classified as Class
0/I (\S~\ref{sres2mass}), and with faint emission at 7~mm. The 7~mm
emission has a secondary peak toward the south, in the same direction
as the extension appreciated in the 1.3~cm emission. The 7~mm emission
does not seem to have contribution from thermal dust emission;
therefore, the centimeter emission, up to 7~mm, is most likely tracing
an ionized wind, as suggested by the spectral index of 0.2--0.6.  The
dust temperature derived from the SED, 26--29~K, is similar to the
median value of the survey, and the associated mass is the smallest,
around 8--12~\mo.

The extended centimeter emission to the northwest consists of a curved
structure with different subcomponents, VLA~2 to VLA~6, and is not
overlapping with the millimeter emission, but starts when the
millimeter emission finishes. This is reminiscent of a bright-rimmed
cloud, as the centimeter emission follows the border of the dust
cloud. We searched the catalog of \citet{reed2003} for nearby OB stars
to the northwest of the millimeter emission, and the closest star was
a B2.5 star at $\sim2.5\degr$ to the north-west (70~pc), which is too
far away to account for the flux of VLA~2 to VLA~6, following
\citet{lefloch1997}. Given the double lobe morphology and the negative
spectral indices of some of these sources (see Fig.~\ref{fi19045_1}
and Table~\ref{tspindex}), they could be tracing extragalactic
objects. However, there are no sources detected at 21~cm in the NVSS,
and we estimated a lower limit for the spectral index of VLA~2 to
VLA~6 between 21 and 3.6~cm of $>-0.03$ (see Table~\ref{tspindex}),
typical of thermal emission. Thus, we favor the interpretation that
these sources are thermal, and have not been detected at 1.3~cm
because their emission is extended and thus (partially) resolved out
by the interferometer.

%---------------------------------------------------------------------
%---------------------------------------------------------------------

%---------------------------------------------------------------------
\begin{figure}[htbp!]
\begin{center}
\begin{tabular}[b]{c}
\noalign{\bigskip}
	\epsfig{file=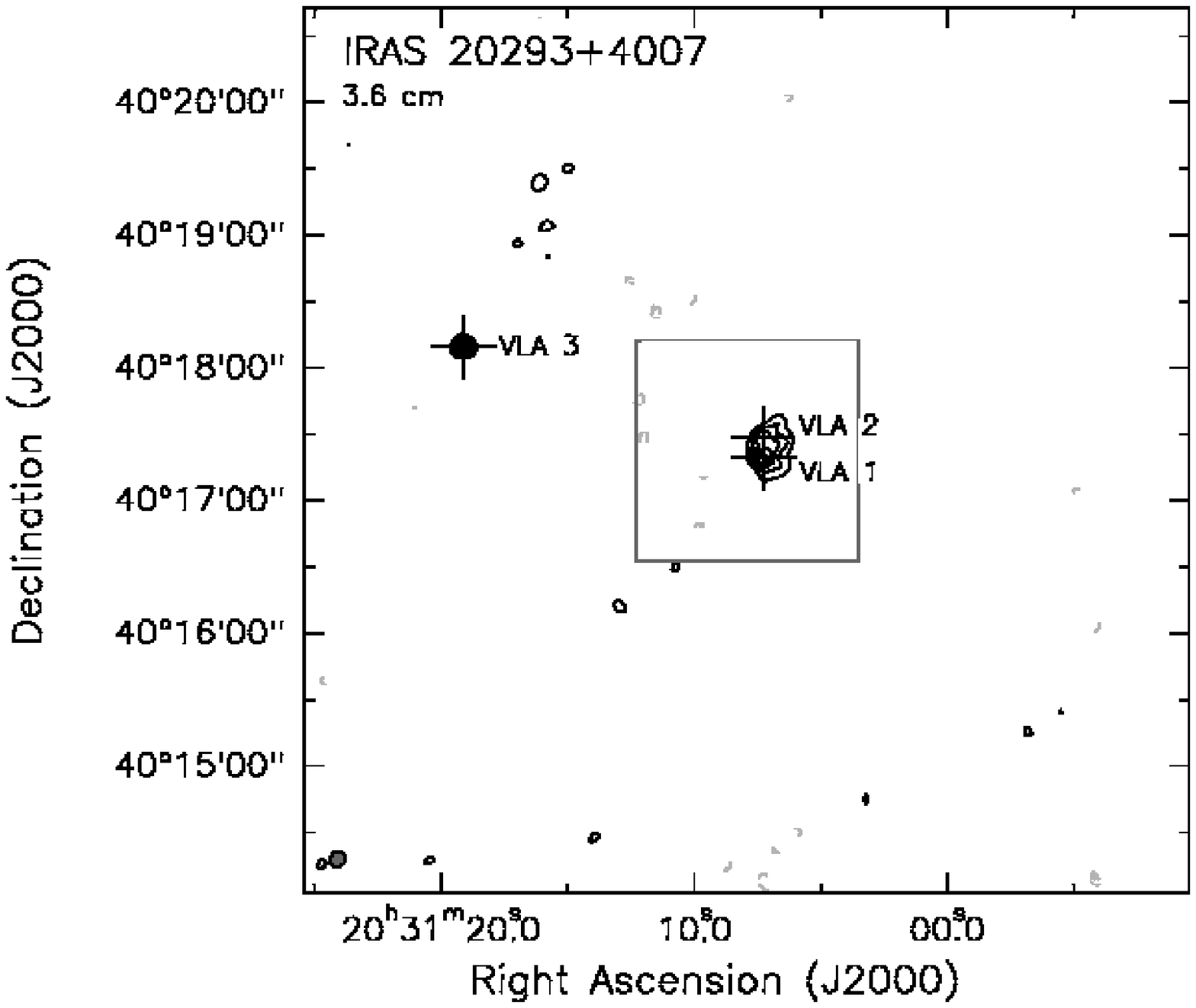, scale=0.48} \\
\noalign{\bigskip}
	\epsfig{file=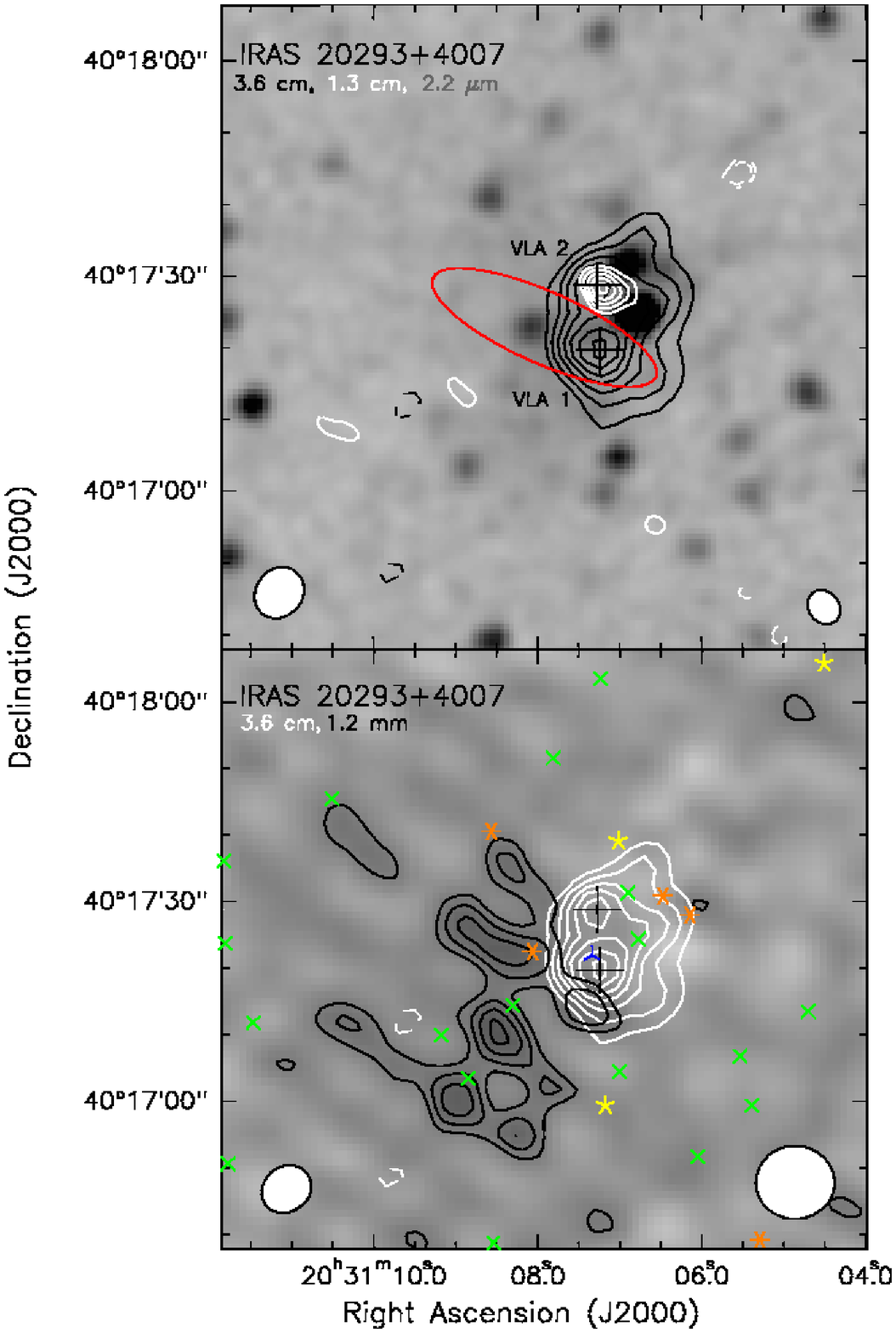, scale=0.48} \\
\noalign{\bigskip}
\end{tabular}
\caption{{\small IRAS~20293+4007.
\emph{Top}: VLA 3.6~cm map. The
box shows the region zoomed in the other panels.
\emph{Middle}: Black contours: VLA 3.6~cm continuum emission. Levels are $-5$,
$-3$ and 3 to 15 in steps of 2, times 0.058~mJy~beam$^{-1}$. White contours: VLA
1.3~cm continuum emission. Levels are $-4$, $-3$, and 3 to 7 in steps of 1,
times 0.080~mJy~beam$^{-1}$. Grey scale: K-band 2MASS image.
\emph{Bottom}: Grey scale and black contours: IRAM~30\,m continuum emission at
1.2~mm. Levels are 3 to 6 in steps of 1,
times 9~mJy~beam$^{-1}$. White contours: 3.6~cm
emission as in middle panel. Color
stars indicate 2MASS sources and white filled symbols refer to masers
(see Fig.~\ref{fi00117_1} for more details).
}}
\label{fi20293_1}
\end{center}
\end{figure}
%---------------------------------------------------------------------

\subsubsection{IRAS~20293+4007\label{i20293}}

At 3.6~cm, I20293 is dominated by one strong source (see
Fig.~\ref{fi20293_1}).  This source has two subcomponents, VLA~1
(south) and VLA~2 (north), and the centimeter peaks are not clearly
associated with any 2MASS source, although a few fall within the
extended centimeter emission. At 1.3~cm there is one strong source
slightly resolved at the position of VLA~2. The spectral index
suggests that VLA~2 is a thermal source, and is likely tracing the
\uchii\ region associated with the IRAS source. However, the IRAS
source could be also associated with a 2MASS source showing infrared
excess that falls just at the eastern edge of the centimeter emission.
The emission at 1.2~mm is extended and clumpy, and shifted to the east
with respect to the centimeter emission (similarly to I00117). There
is no millimeter emission directly associated with the centimeter
emission, suggesting that the \uchii\ region (possibly VLA~2) has
disrupted and pushed the surrounding gas away. This could be again the
reason why \nh\ is not detected in the region \citep{molinari1996}.

%---------------------------------------------------------------------
\begin{figure*}[t!]
\begin{center}
\begin{tabular}[b]{cc}
\noalign{\bigskip}
	\epsfig{file=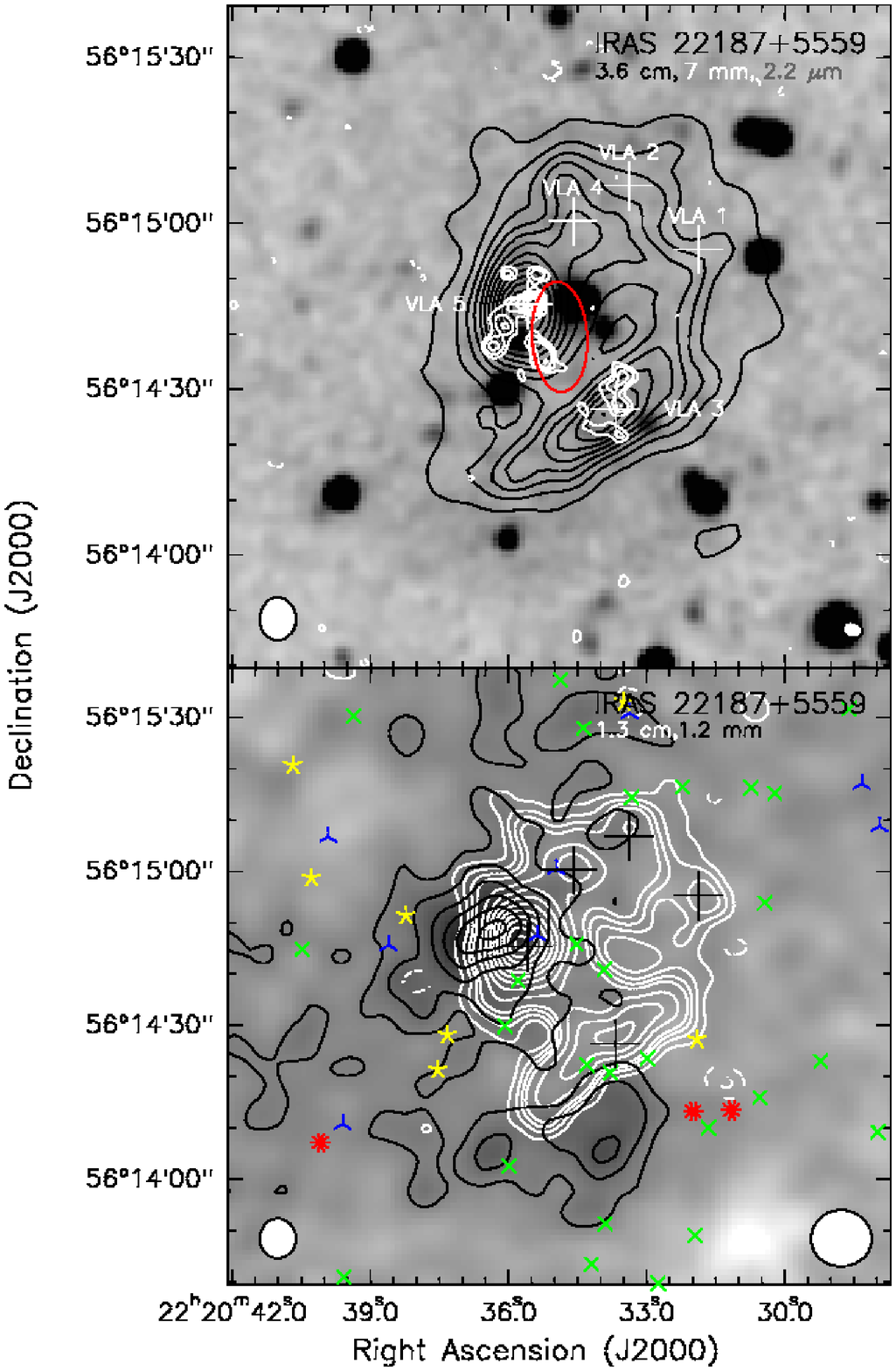, scale=0.49} &
	\epsfig{file=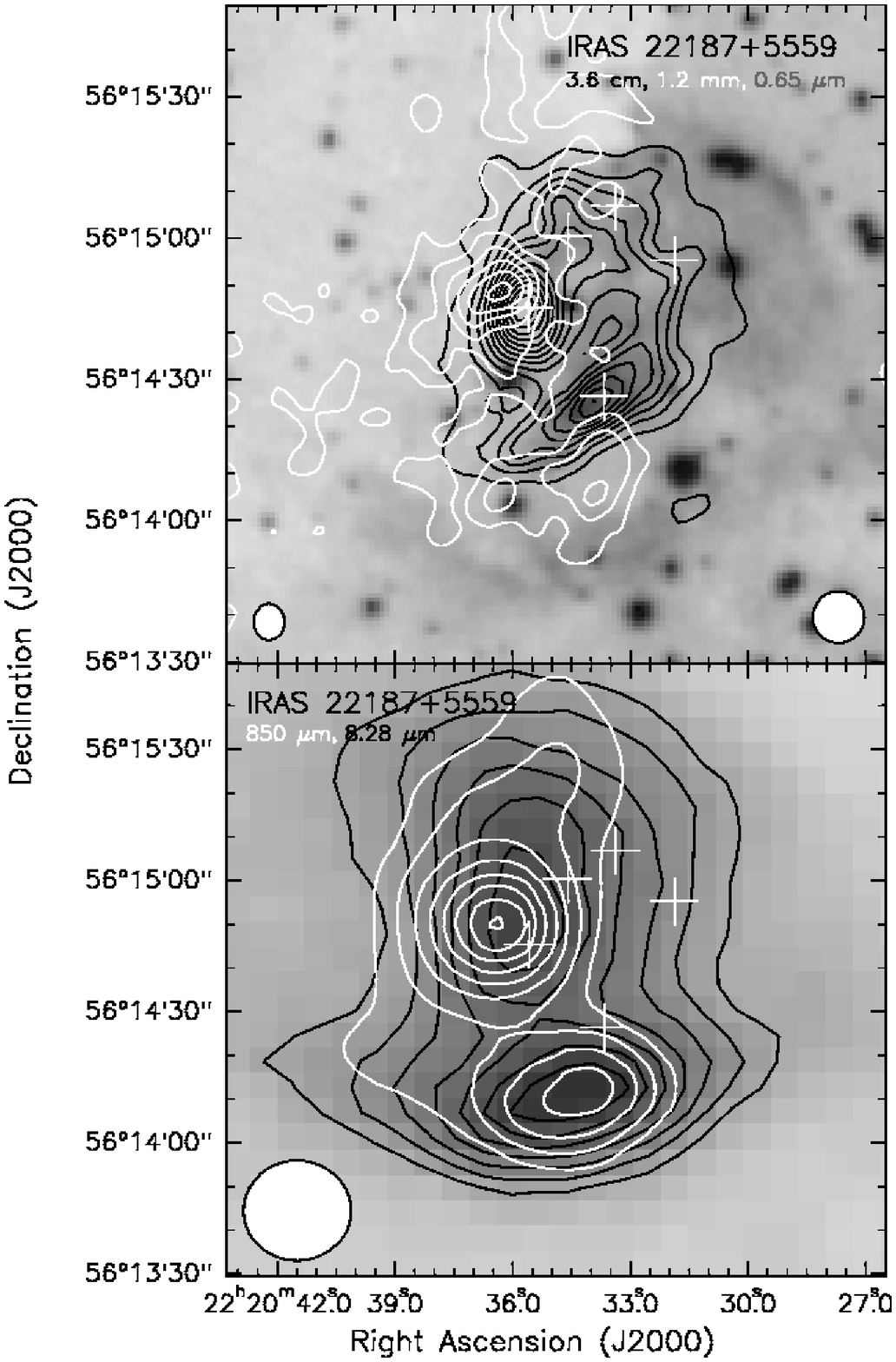, scale=0.49} \\
\noalign{\bigskip}
\end{tabular}
\caption{{\small IRAS~22187+5559.
\emph{Top-left}: Black contours: VLA 3.6~cm continuum emission. Levels are $-5$,
$-3$ and 3 to 42 in steps of 3, times 0.155~mJy~beam$^{-1}$. White contours: VLA
7~mm continuum emission. Levels are $-4$, $-3$, and 3 to 9 in steps of 1, times
0.178~mJy~beam$^{-1}$. Grey scale: K-band 2MASS image.
\emph{Bottom-left}: Grey scale and black contours: IRAM~30\,m continuum emission
at 1.2~mm. Levels are 2 to 12 in steps of 2, times 7~mJy~beam$^{-1}$. White
contours: VLA 1.3~cm continuum emission. Levels are $-5$, $-3$, 3, 5, 7 and 12
to 75 in steps of 7, times 0.060~mJy~beam$^{-1}$.
\emph{Top-right}: Grey scale: POSS-II red image. Black contours: 3.6~cm emission
as in top-left panel. White contours: 1.2~mm emission as in bottom-left panel .
\emph{Bottom-right}: Grey scale and black contours: A-band MSX image
(8.28~$\mu$m). White contours: SCUBA 850~$\mu$m emission. Levels are 3 to 17 in
steps of 2, times 0.045~Jy~beam$^{-1}$ \citep{difrancesco2008}. 
Color
stars indicate 2MASS sources and white filled symbols refer to masers
(see Fig.~\ref{fi00117_1} for more details).
}}
\label{fi22187_1}
\end{center}
\end{figure*}
%---------------------------------------------------------------------

\subsubsection{IRAS~22187+5559\label{i22187}}

I22187 contains the most extended centimeter source of this survey
(see Fig.~\ref{fi22187_1}). The morphology of the centimeter emission
reveals two main subcomponents, encompassed by a common structure
extending toward the northwest. The strongest subcomponent, VLA~5,
located in the north, is compact, has an extension toward the
northwest and could be associated with a 2MASS source (with no
infrared excess) lying at about $3''$ to the northwest from the
centimeter peak. The second subcomponent, VLA~3, fainter and located
in the south, is elongated in the southeast-northwest direction. At
1.3~cm, the emission strongly resembles the 3.6~cm emission, with
three faint additional components within the extended emission, VLA~1
to the north of VLA~3, and VLA~2 and VLA~4 to the northwest of VLA~5.
The spectral index of VLA~3 and VLA~5 is flat for both sources, while
for the three faint sources, the spectral index varies from flat to
$+0.6$. These three faint sources could be globules externally ionized
by VLA~3 and VLA~5. \citet{garay2006} find in IRAS~16128$-$5109 a
similar case of centimeter emission coming from externally ionized
globules.

The large-scale 1.2~mm emission also presents two main condensations,
which are slightly ($\sim\!7''$) displaced from VLA~5 (millimeter peak
to the east) and VLA~3 (millimeter peak to the south). The 1.2~mm
emission is likely associated with the HCO$^+$ emission
\citep{richards1987} and the CO emission detected in the region
\citep{wouterlootbrand1989}. SCUBA maps at 850 and 450~$\mu$m
\citep{difrancesco2008} show condensations associated with the 1.2~mm
peaks (see Fig.~\ref{fi22187_1}). We estimated a flux density of
$0.7\pm0.2$~Jy at 850~$\mu$m and $6.1\pm2.2$~Jy at 450~$\mu$m for the
condensation associated with VLA~3, and a flux density of
$1.9\pm0.2$~Jy at 850~$\mu$m and $12.0\pm2.9$~Jy at 450~$\mu$m for the
condensation associated with VLA~5. In the mid-infrared there are two
MSX sources whose morphology and position are very similar to the two
1.2~mm, 850~$\mu$m, and 450~$\mu$m condensations. In addition, the
POSSII red plates reveal a zone of high extinction starting at the
position of VLA~5 and extending to the east, which matches with the
MSX and 1.2~mm sources (see Fig.~\ref{fi22187_1}).  Another zone of
high extinction is also found to the south of VLA~3 (see
Fig.~\ref{fi22187_1}). These high extinction zones, traced by the
millimeter and mid-infrared emissions, are likely tracing a dense
molecular cloud seen against a bright background. We note that while
VLA~3 is associated with optical emission, VLA~5 lies within the
molecular cloud (dark in the optical), and thus seems to be more
embedded and probably younger than VLA~3. Note also that the extension
of the centimeter emission toward the west of VLA~5 and VLA~3 is
suggestive of the material surrounding the centimeter peaks having a
lower density on the west side than on the east/southern sides, where
the higher-density gas could be slowing down the expansion of the
ionized gas.

A cluster of 7~mm sources is associated with both VLA~5 and VLA~3. The
strongest 7~mm source, associated with VLA~5, is elongated in the
east-west direction, surrounded by different fainter 7~mm sources, and
possibly associated with the closest 2MASS source, which is shifted
only 2$\farcs$1. For VLA~3, the emission has several components with
the strongest one located $\sim\,5''$ to the north of the centimeter
peak. All the 7~mm emission associated with VLA~3 and VLA~5 comes from
thermal free-free emission. We built and fitted two different SEDs for
VLA~3 and VLA~5, and obtained similar temperatures for both sources.
The associated mass is higher for VLA~5 ($\sim\!35$~\mo) than for
VLA~3 ($\sim\!15$~\mo).

\subsubsection{IRAS~22198+6336\label{i22198}}

This source is located in the L1204/S140 molecular complex.
\citet{tafalla1993}, using the CS(1--0) transition, detect seven cores
in a region of $\sim\!50'$, named A to G. L1204-G is associated with
I22198. (Note that some authors name I22198 as L1204-A, which has the
coordinates of IRAS~22176+6303, so one must be cautious when using the
labeling of \citealt{tafalla1993}). Observations of CO and C$^{18}$O
by \citet{jenness1995} and observations of CO by \citet{zhang2005}
show evidences of molecular outflow emission. There is also H$_2$O
maser emission detected by \citet{palla1991}, \citet{valdettaro2002}
and \citet{felli2007}, who find several bursts of variable duration
(200--500~days).

\begin{figure}[htbp!]
\begin{center}
\begin{tabular}[b]{c}
\noalign{\bigskip}
	\epsfig{file=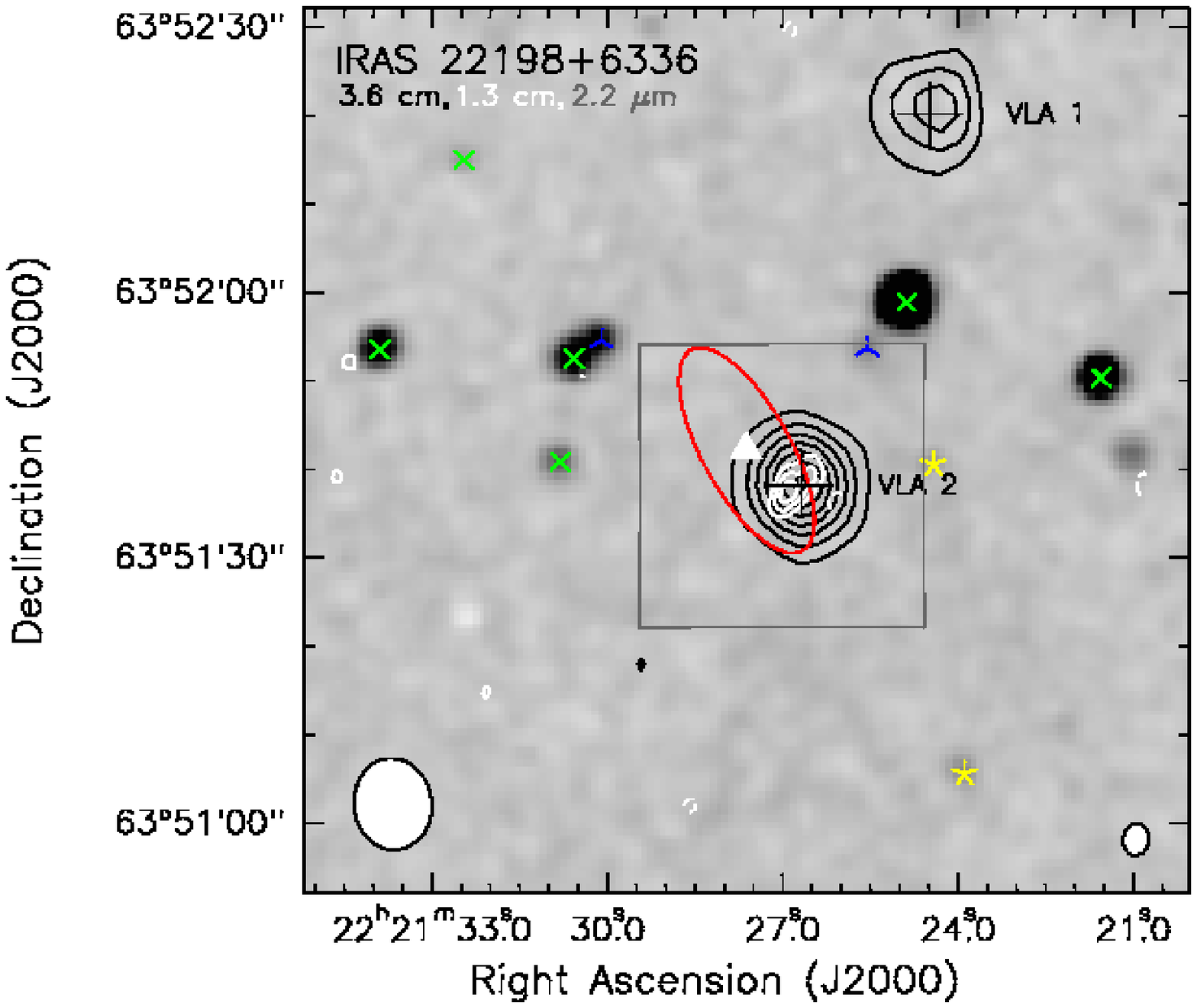, scale=0.49} \\
\noalign{\bigskip}
	\epsfig{file=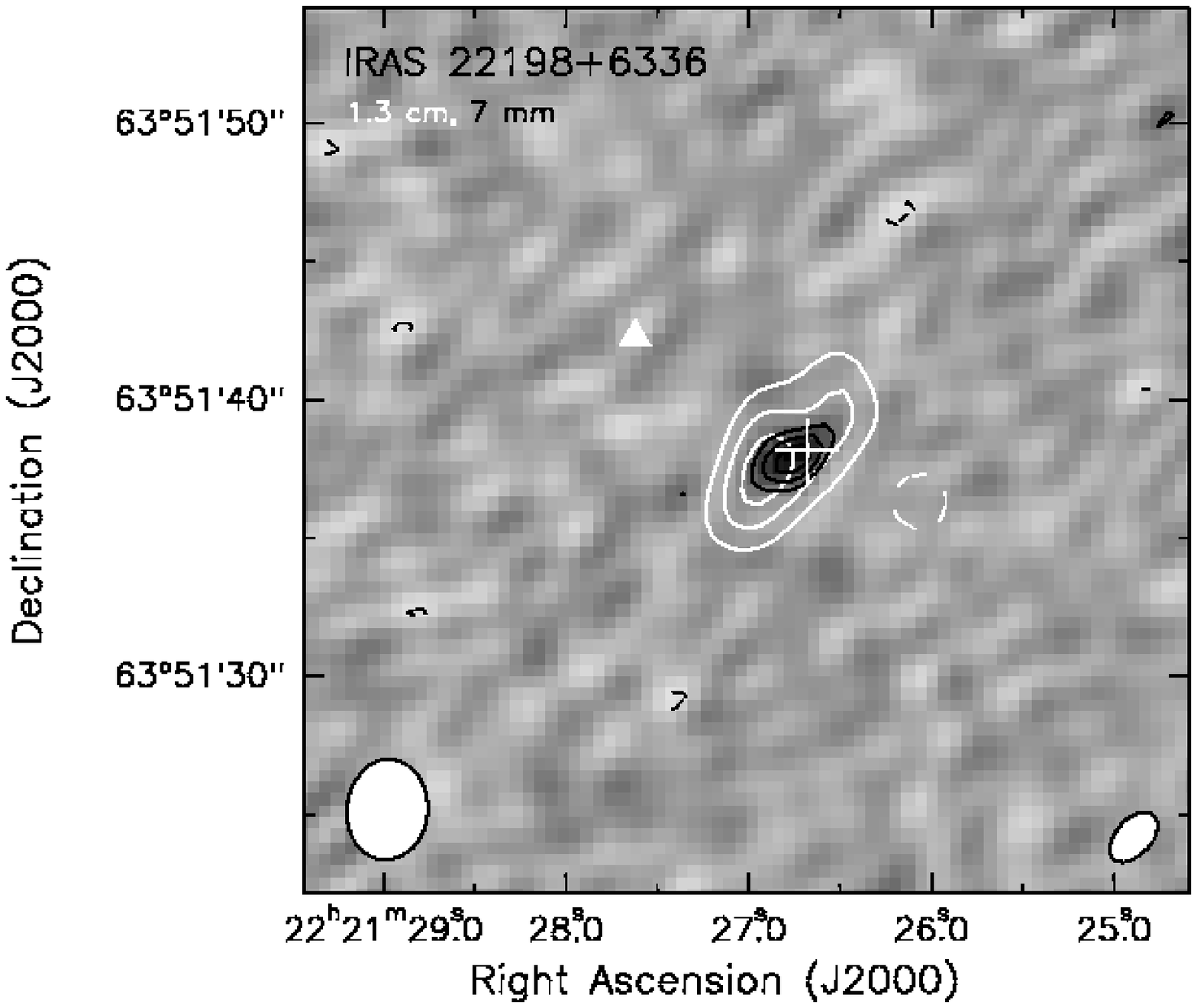, scale=0.49} \\
\noalign{\bigskip}
\end{tabular}
\caption{{\small IRAS~22198+6336.
\emph{Top}: Black contours: VLA 3.6~cm continuum emission. Levels are
$-6$, $-3$ and 3 to 21 in steps of 3, times 0.027~mJy~beam$^{-1}$.
White contours: VLA 1.3~cm continuum emission. Levels are $-5$, $-3$,
and 3 to 9 in steps of 2, times 0.063~mJy~beam$^{-1}$. Grey scale:
K-band 2MASS image. The box shows the region zoomed in the bottom
panel.
\emph{Bottom}: Grey scale and black contours: VLA 7~mm continuum emission.
Levels are $-5$, $-3$, and 3 to 7 in steps of 2, times 0.226~mJy~beam$^{-1}$.
White contours: 1.3~cm emission as in top panel. Color
stars indicate 2MASS sources and white filled symbols refer to masers
(see Fig.~\ref{fi00117_1} for more details).
}}
\label{fi22198_1}
\end{center}
\end{figure}

Emission at 3.6~cm shows a compact single source lying within the IRAS
error ellipse (see Fig.~\ref{fi22198_1}), with a counterpart at
1.3~cm, but with no 2MASS neither MSX sources associated. The
centimeter peaks spatially coincide ($\sim\!5''$ to the northeast)
with a strong and compact submillimeter condensation detected at 450
and 850~\mum\ by \citet{jenness1995}. In addition, emission at 7~mm
has been detected toward the source, with 44\% of the flux at 7~mm
coming from thermal dust emission. The mass of the 7~mm dust component
is $\sim\!9$~\mo. The source at 1.3~cm and 7~mm is elongated in the
southeast-northwest direction (see Fig.~\ref{fi22198_1}~bottom panel),
with a structure reminiscent of a radiojet, consistent with the
spectral index measured in the centimeter range, $\sim\!0.5$.
Furthermore, the large-scale outflow found in this region is slightly
elongated in the southeast-northwest direction. However, since a
substantial part of the 7~mm flux density is coming from thermal dust
emission, at this wavelength we are probably observing the
superposition of an ionized wind and a dust disk. If the direction of
the outflow is the same as the direction of the elongation of the
1.3~cm and 7~mm emissions, one could infer from the measured
deconvolved size of the 7~mm source, $3250\times260$~AU, an upper
limit for the size of the disk, $\sim\!300$~AU, as the disk is
expected to be perpendicular to the 7~mm source.  It would be
necessary to study the outflow with high angular resolution to confirm
its direction and thus the upper limit for the size of the disk. It is
worth noting that as no near/mid-infrared sources are associated with
the centimeter emission (note that there is no distance effect because
this is the source of the survey with the shortest distance), this
source must be deeply embedded in cold dust (as indicated by the
derived dust temperature of 25--29~K). We propose that VLA~2 is the
youngest source of our survey and is tracing an early-type B protostar
in an evolutionary stage similar to Class 0 sources in the low-mass
regime.

%---------------------------------------------------------------------
%---------------------------------------------------------------------

%---------------------------------------------------------------------
\begin{figure}[htbp!]
\begin{center}
\begin{tabular}[b]{c}
\noalign{\bigskip}
	\epsfig{file=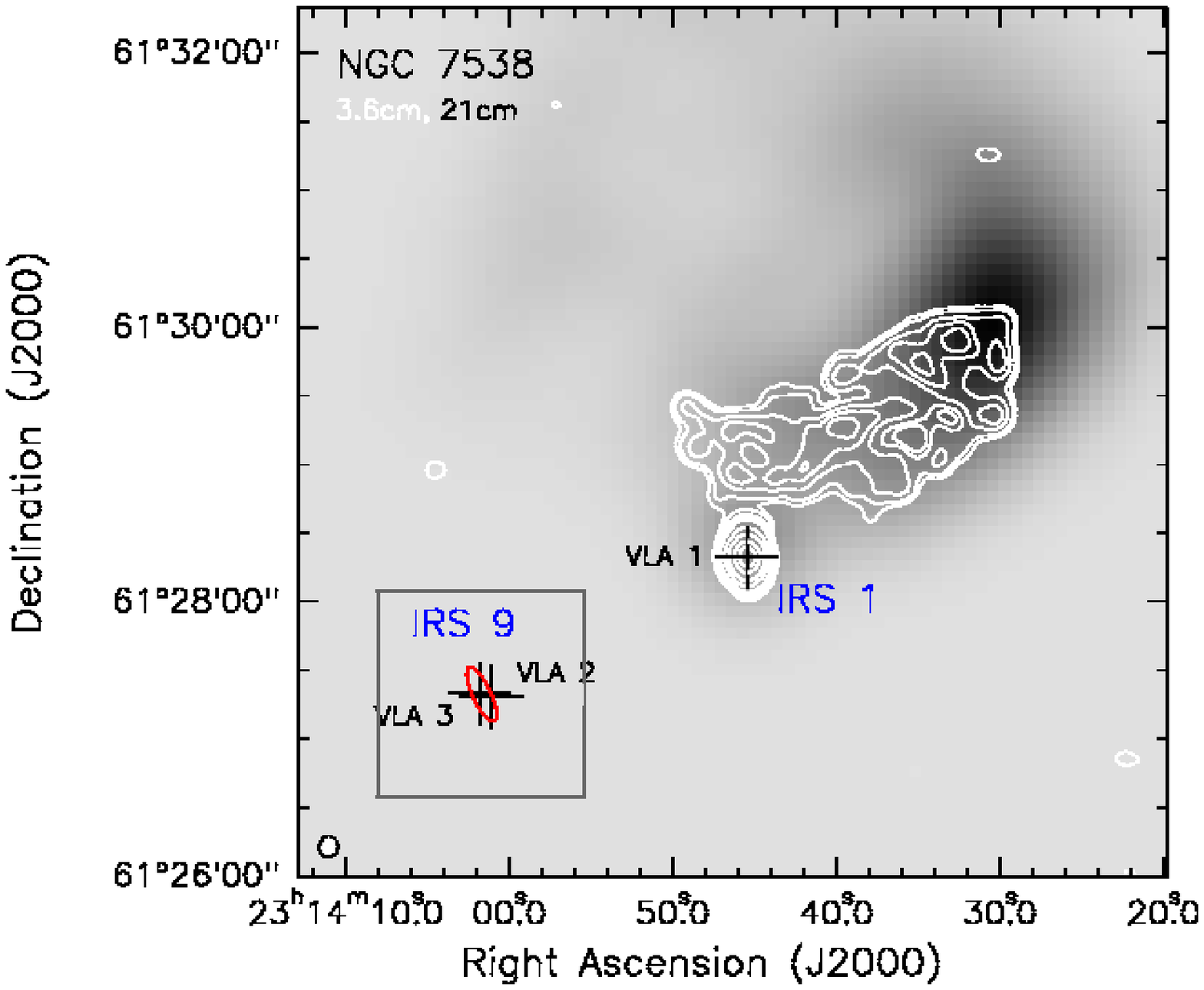, scale=0.48} \\
\noalign{\bigskip}
	\epsfig{file=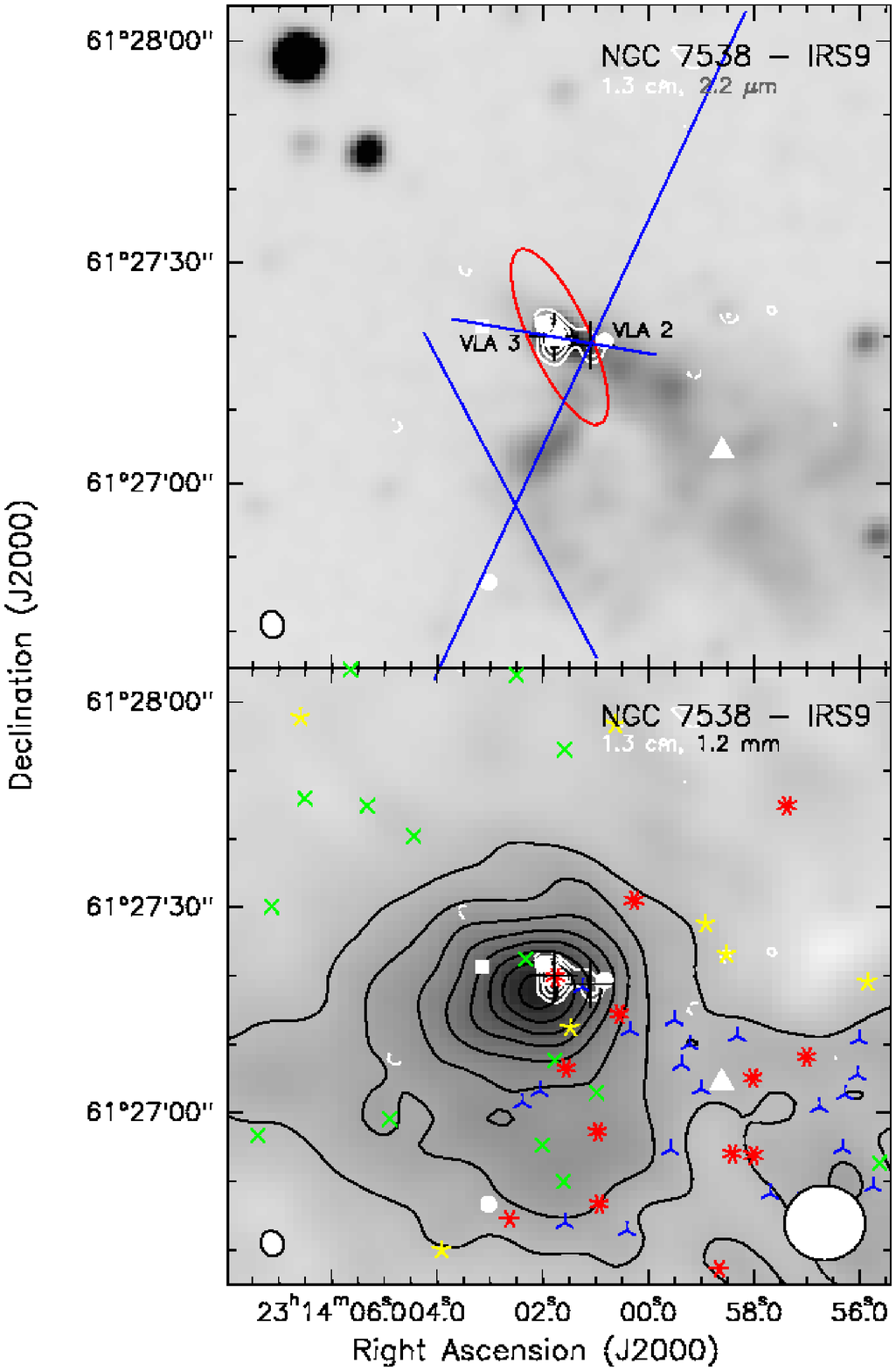, scale=0.48} \\
\noalign{\bigskip}
\end{tabular}
\caption{{\small NGC\,7538-IRS9.
\emph{Top}: Grey scale: 21~cm emission from NVSS. White contours: VLA
3.6~cm continuum emission. Levels are 3, 5, 10, 15, 20, 50, 100, 200,
300 and 400, times 1.28~mJy~beam$^{-1}$. The box shows the region
zoomed in the other panels.
\emph{Middle}: White contours: VLA 1.3~cm continuum emission. Levels are $-5$,
$-3$ and 3 to 11 in steps of 2, times 0.204~mJy~beam$^{-1}$. Grey scale: K-band
2MASS image. Blue lines correspond to molecular outflows \citep{sandell2005}.
\emph{Bottom}: Grey scale and black contours: IRAM~30\,m continuum emission at
1.3~mm. Levels are 5 to 35 in steps of 3, times 30~mJy~beam$^{-1}$
\citep{sandellsievers2004}.
White contours: 1.3~cm emission as in middle panel. Color
stars indicate 2MASS sources and white filled symbols refer to masers
(see Fig.~\ref{fi00117_1} for more details).
}}
\label{fi23118_1}
\end{center}
\end{figure}
%---------------------------------------------------------------------

%---------------------------------------------------------------------
\begin{figure}[ht!]
\begin{center}
\begin{tabular}[b]{c}
\noalign{\bigskip}
	\epsfig{file=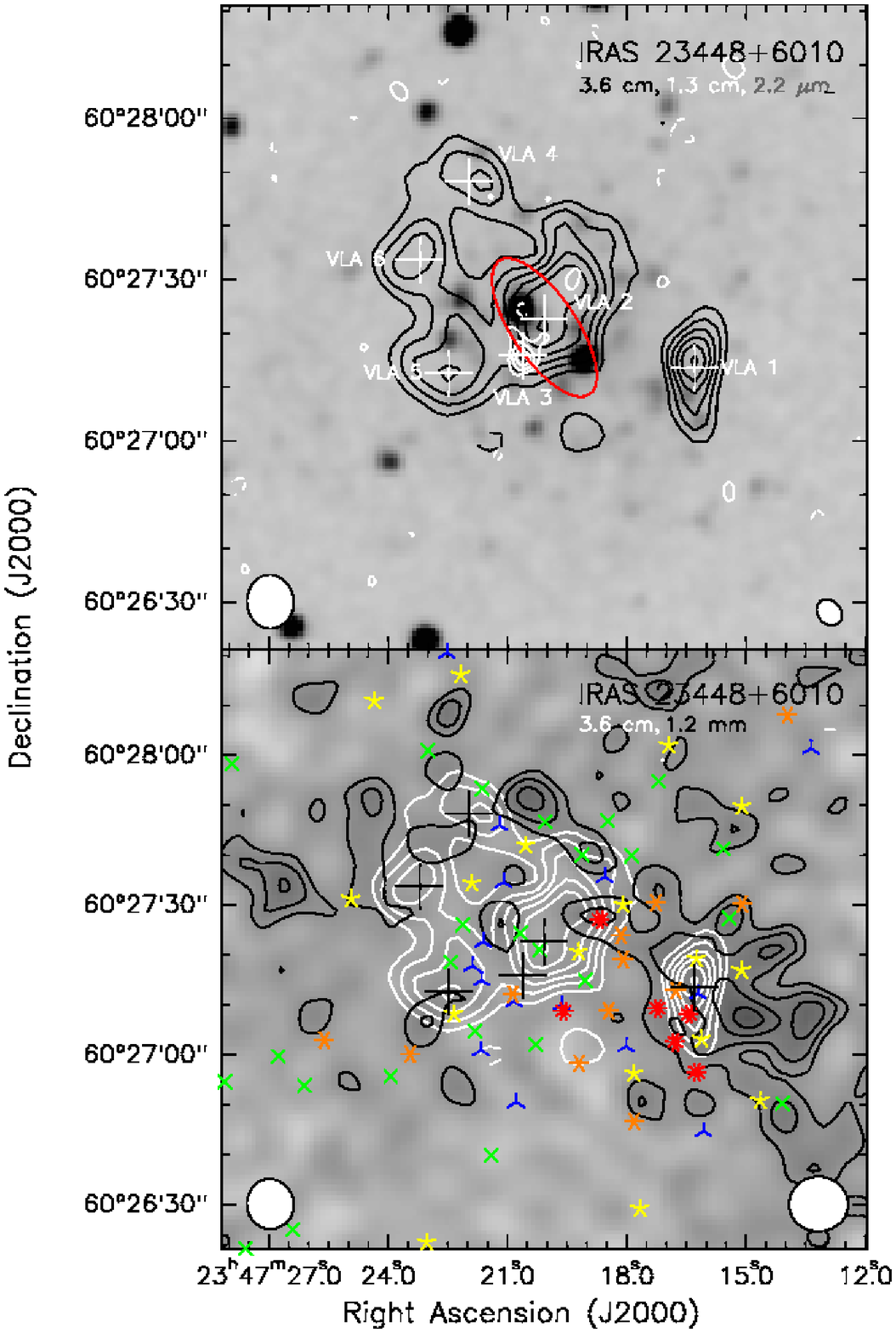, scale=0.49} \\
\noalign{\bigskip}
\end{tabular}
\caption{{\small IRAS~23448+6010.
\emph{Top}: Black contours: VLA 3.6~cm continuum emission. Levels are $-5$, $-3$
and 3 to 13 in steps of 2, times 0.030~mJy~beam$^{-1}$. White contours: VLA
1.3~cm continuum emission. Levels are $-5$, $-3$, 3, 5, and 7, times
0.076~mJy~beam$^{-1}$. Grey scale: K-band 2MASS image.
\emph{Bottom}: Grey scale and black contours: IRAM~30\,m continuum emission at
1.2~mm. Levels are 2 to 5 in steps of 1, times 11~mJy~beam$^{-1}$. White contours: 3.6~cm emission
as in bottom-left panel. Color
stars indicate 2MASS sources and white filled symbols refer to masers
(see Fig.~\ref{fi00117_1} for more details).
}}
\label{fi23448_1}
\end{center}
\end{figure}
%---------------------------------------------------------------------

\subsubsection{NGC\,7538-IRS9\label{i23318}}

We detected one source at 1.3~cm (at 3.6~cm we were limited by dynamic
range), associated with the IRAS source and with a 2MASS source at the
northeastern border of an infrared nebulosity (see
Fig.~\ref{fi23118_1}). The source at 1.3~cm splits up into two
subcomponents, VLA~2 and VLA~3, being VLA~3 clearly associated with
the infrared source IRS~9.

\nh\ observations were carried out by \citet{zheng2001}, and CO and
H$_2$ observations by \citet{davis1998}, who discover a H$_2$ jet in
the southeast-northwest direction. \citet{sandell2005} detect three
molecular outflows in the region: one HCO$^+$ high-velocity outflow
associated with IRS~9 (VLA~3); another outflow associated with a
protostellar core located $\sim\!25''$ to the southeast of IRS~9; and
a third outflow, detected in the CO\,(1--0) transition, coincident
with the H$_2$ jet, whose driving source is traced by an infrared peak
labeled "A1" by \citet{tamura1991}. This infrared peak coincides with
the position of the VLA~2 peak. Thus, the centimeter source detected
is probably the counterpart of the infrared source driving the H$_{2}$
jet \citep{sandell2005}. H$_2$O, OH, and CH$_3$OH maser emission has
been detected \citep{kameya1990,sandell2005} in IRS~9 and in the
outflows.

The large-scale millimeter emission toward NGC\,7538 has been studied
by a number of authors. \citet{reidwilson2005} map the submillimeter
continuum emission at 450 and 850~$\mu$m, and detect a single source
at  both wavelengths, clearly associated with IRS~9, with a faint
extension to the south likely associated with the infrared nebulosity.
\citet{sandellsievers2004} map the continuum emission at 350~$\mu$m
and 1.3~mm, finding similar results.

The region was observed at 7~mm by \citet{vandertakmenten2005} with
the VLA in the C and A configurations. These authors detect one source
at the position of VLA~3 (positions agree within $0\farcs3$), which is
elongated in the north-south direction in the A-configuration
observations. This is consistent with the deconvolved position angle
that we have obtained toward VLA~3, $10\degr$. The present
observations at 1.3~cm add a new point to the centimeter SED, which
was fitted through an ionized accretion flow model by
\citet{vandertakmenten2005}. Our value at 1.3~cm is about a factor of
2 higher than that predicted by the model \citep[see Fig.~4
of][]{vandertakmenten2005}, most likely because our observations were
carried out in the D configuration and thus picked up more extended
emission. The dust temperature from the SED fitted by us (which 
includes both VLA~2 and VLA~3) using a modified blackbody law,  is
near the median value of the survey, and the associated mass,
50--80~\mo, is among the highest in the survey. The small deconvolved
size of the 1.3~cm source, its association with a 7~mm source
\citep{vandertakmenten2005} and with the peak of a compact 1.3~mm
source \citep{sandellsievers2004}, the high infrared excess of the
2MASS source associated with VLA~3, together with the fact that the
centimeter fluxes are consistent with an ionized accretion flow
\citep{vandertakmenten2005} suggest that VLA~3 (and probably VLA~2) is
among the youngest sources in the survey.

\subsubsection{IRAS~23448+6010\label{i23448}}

The 3.6~cm emission toward this region shows a number of sources that
lie in a region with a cluster of 2MASS sources showing strong
infrared excess \citep[][see Fig.~\ref{fi23448_1}]{kumar2006}. The
1.2~mm emission of the region is faint and extended and surrounds the
northern and western edges of the centimeter emission. Three of the
centimeter sources, VLA~1 to VLA~3, have spectral indices compatible
with thermal free-free optically thin emission, and thus could be
\uchii\ regions, although with different properties: while VLA~3 is
the only centimeter source detected at 1.3~cm, and has no 2MASS
counterpart, VLA~1 and VLA~2 are only detected at 3.6~cm and are
clearly associated with 2MASS sources, with VLA~2 falling inside the
IRAS error ellipse. As the 2MASS source coincident with VLA~2 is not
detected in the $J$-band, we estimated its infrared excess from the
$(H-K)$ color alone, which is 0.63, typical of T-Tauri stars
\citep[\eg][]{kenyon1995}, and thus included in the group of sources
with infrared excess $<0.4$ (following \S~\ref{sres2mass}). We note
that the centimeter emission of VLA~2 to VLA~6 is reminiscent of a
ring-like structure and that this emission could be produced by the
ionizing photons from the \uchii\ regions. However, none of these
sources lie at the center of the ring, and in addition, the rate of
ionizing photons coming from VLA~2 or VLA~3 is too low (more than one
order of magnitude) to produce the centimeter emission of VLA~4, VLA~5
and VLA~6 \citep[following][]{lefloch1997}. Thus, the centimeter
emission of I23448 may be tracing a young cluster of B-type stars,
which could be dispersing the surrounding material and would be
responsible for the morphology of the millimeter emission.
\citet{garay2006} find a similar case of a cluster of centimeter
sources in IRAS~15502$-$5302, but with the millimeter emission still
associated with the centimeter sources. HCO$^+$ and HCN emission has
been detected toward I23448 \citep{richards1987} as well as CO
emission \citep{wouterlootbrand1989}.

%---------------------------------------------------------------------
\begin{figure*}[ht!]
\begin{center}
\begin{tabular}[b]{l}
\noalign{\bigskip}
	\epsfig{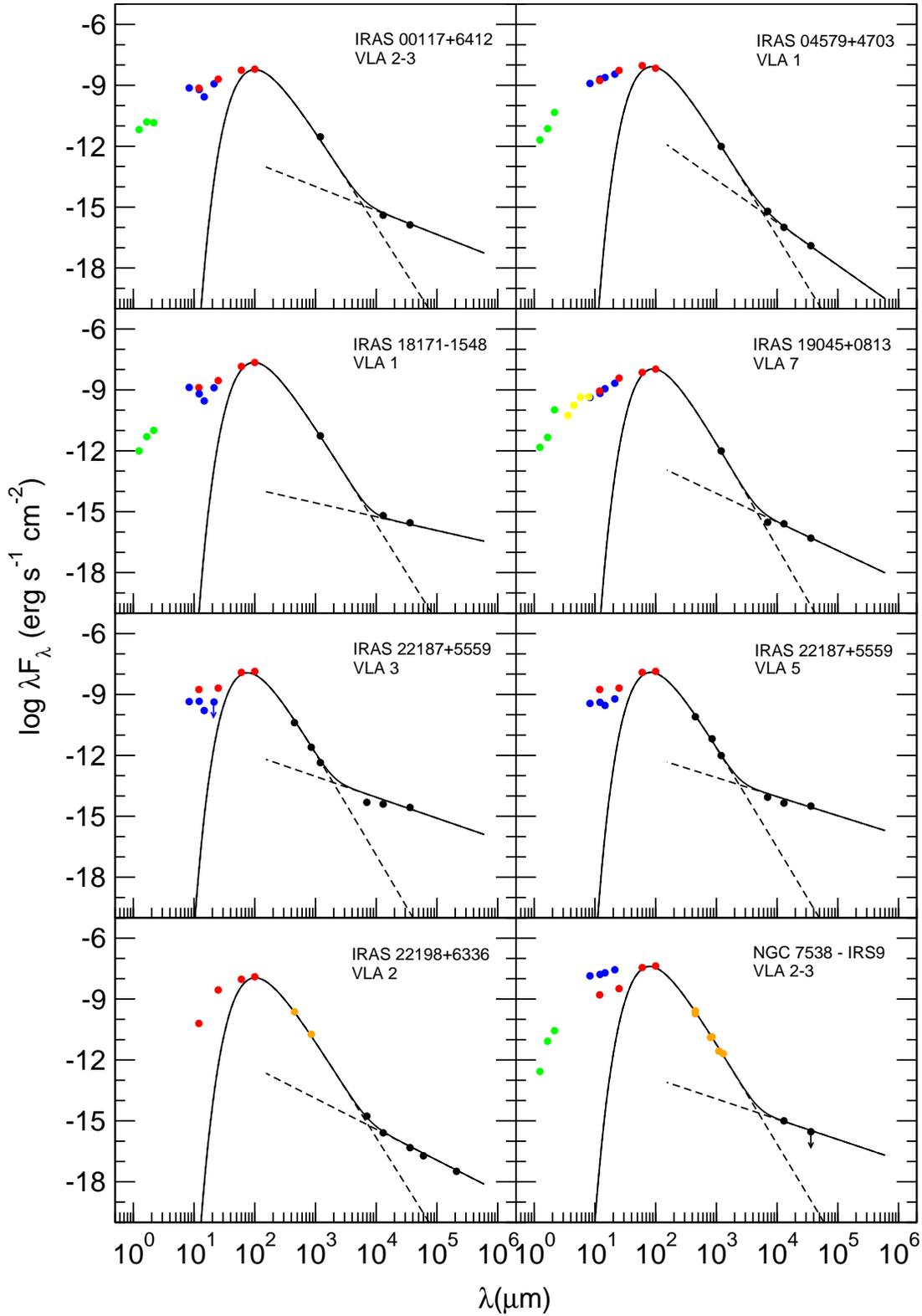} \\
\noalign{\bigskip}
\end{tabular}
\caption{{\small
Spectral Energy Distributions. Green dots: 2MASS data. Yellow dots:
IRAC-\emph{Spitzer} data. Blue dots: MSX data. Red dots: IRAS data.
Orange dots: millimeter and submillimeter data from \citet{jenness1995}
for IRAS~22198+6336, and from \citet{sandellsievers2004} for
NGC\,7538-IRS9. Black dots: this work, and submillimeter data from
SCUBA (JCMT) for IRAS~22187+5559. Dashed lines: modified blackbody and
free-free optically thin fits for the far-infrared to centimeter
wavelengths. Black line: sum of the modified blackbody law and the
free-free optically thin law. The results of the fit are shown in
Table~\ref{tsed}. Note that there is no attempt to fit the near- and
mid-infrared wavelengths.
}}
\label{fi_sed}
\end{center}
\end{figure*}
%---------------------------------------------------------------------

%-------------------------------------------------------------------------------
\begin{table*}
\caption{Evolutionary indicators of the sources, classified from smallest to largest disruption degree (see Section~\ref{sdis})}
\begin{center}
\scriptsize{
\begin{tabular}{llccccccccccc}
\hline\hline\noalign{\smallskip}
&VLA
&cm
&cm-mm
&mm
&Disruption
&Dense
&
&\multicolumn{3}{c}{Maser Emission}
&2MASS
&MSX
\\
\cline{9-11}
\noalign{\smallskip}
Region
&Source
&Structure
&Offset
&Structure
&Degree$^\mathrm{a}$
&Gas
&Outflow
&\chtoh\
&\water\
&OH
&IR Excess
&Emission
\\
\noalign{\smallskip}
&
&
&(pc)
&
&(\%)
&
&
&
&
&
&
&
\\
%\noalign{\smallskip}
\noalign{\smallskip}
\hline\noalign{\smallskip}
IRAS~22198+6336		&3
	&disk / radiojet	&$\lesssim\!0.03$	&compact$^\mathrm{b}$ &\phn7$^\mathrm{b}$		&\nh, CS	&yes\phe	& &yes &no		&\ldots$^\mathrm{c}$	&no	\\

\noalign{\smallskip}
\hline\noalign{\smallskip}
IRAS~19045+0813		&7
	&compact		&$\lesssim\!0.04$	&compact\phe &46\phe		&CS		&	&no &yes &no		& orange\phe		&yes	\\

\noalign{\smallskip}
\hline\noalign{\smallskip}
IRAS~04579+4703		&1
	&compact		&$\lesssim\!0.06$	&compact\phe &56\phe		&\nh		&yes$^\mathrm{d}$	&no &yes &no		& red\phe	&yes	\\

\noalign{\smallskip}
\hline\noalign{\smallskip}
NGC\,7538-IRS9		&3
	&compact		&$\sim\!0.09$		&compact$^\mathrm{b}$ &69$^\mathrm{b}$		&\nh, HCN	&yes\phe	&yes &yes &yes		& red\phe		&yes	\\

\noalign{\smallskip}
\hline\noalign{\smallskip}
IRAS~18171$-$1548	&1
	&compact		&$\sim\!0.08$		&peak+ext\phe &78\phe		&no		&	&no &no &no		& yellow\phe		&yes	\\

\noalign{\smallskip}
\hline\noalign{\smallskip}
IRAS~22187+5559		&5
	&peak+ext		&$\sim\!0.10$		&peak+ext\phe &78\phe		&HCO$^+$	&yes$^\mathrm{d}$	& &no &no	& blue$^\mathrm{e}$	&yes	\\

\noalign{\smallskip}
\hline\noalign{\smallskip}
IRAS~00117+6412		&2--3
	&compact		&$\sim\!0.13$		&peak+ext\phe &81\phe		&\nh, CS	&no$^\mathrm{f}$	&no &yes &no	& yellow\phe	&yes	\\

\noalign{\smallskip}
\hline\noalign{\smallskip}
IRAS~23448+6010		&2
	&ext			&$\sim\!0.19$		&dispersed\phe &82\phe		&HCN, HCO$^+$		&	& &no &no		&\ldots$^\mathrm{g}$		&yes		\\

\noalign{\smallskip}
\hline\noalign{\smallskip}
IRAS~22187+5559		&3
	&peak+ext		&$\sim\!0.10$		&peak+ext\phe &87\phe		&HCO$^+$	&yes$^\mathrm{d}$	& &no &no	& green$^\mathrm{e}$	&yes	\\

\noalign{\smallskip}
\hline\noalign{\smallskip}
IRAS~20293+4007		&2
	&compact		&$\sim\!0.25$		&dispersed\phe &91\phe	 	&no		&	& &no &no		& green$^\mathrm{e}$	&yes	\\

\hline
\end{tabular}
\begin{list}{}{}
\item[$^\mathrm{a}$] Defined in Equation~\ref{eq_disruption_degree}. The error of the disruption degree is $\sim\,4\%$.
\item[$^\mathrm{b}$] I22198 850~$\mu$m map from \cite{jenness1995},
and NGC\,7538-IRS9 1.3~mm map from \cite{sandellsievers2004}.
\item[$^\mathrm{c}$] Source not detected in the 2MASS catalog.
\item[$^\mathrm{d}$] Weak outflow emission.
\item[$^\mathrm{e}$] 2MASS sources falling inside the centimeter emission, but slightly offset from the centimeter peak.
\item[$^\mathrm{f}$] Outflow emission detected toward a source embedded in the dust condensation but not toward the centimeter source.
\item[$^\mathrm{g}$] See Section~\ref{i23448}.
\item[Note.-] The values of this table come from this work and from
different references for each source. References for I00117: 4, 6, 14,
17, 28, 29; I04579: 7, 17, 26, 27, 28, 29; I18171: 4, 17, 19, 25;
I19045: 3, 4; I20293: 17, 19; I22187: 7, 9, 19, 20, 28; I22198: 7, 8,
10, 12, 15, 17, 19, 23, 24, 26, 28, 29, 31; NGC\,7538-9: 2, 11, 13, 16,
21, 30; I23448: 5, 7, 18, 19, 20, 27, 28.
\item[References:] 
(1) \citealt{bachiller1990}, 
(2) \citealt{boogert2002}, 
(3) \citealt{brand1994}, 
(4) \citealt{bronfman1996}, 
(5) \citealt{casoli1986}, 
(6) \citealt{cesaroni1988}, 
(7) \citealt{edris2007}, 
(8) \citealt{furuya2003}, 
(9) \citealt{galt1989}, 
(10) \citealt{deGregorio2006}, 
(11) \citealt{hasegawa1995}, 
(12) \citealt{jenness1995}, 
(13) \citealt{kameya1990}, 
(14) \citealt{kimkurtz2006}, 
(15) \citealt{larionov1999}, 
(16) \citealt{mitchell1991}, 
(17) \citealt{molinari1996}, 
(18) \citealt{nguyenqrieu1987}, 
(19) \citealt{palla1991}, 
(20) \citealt{richards1987}, 
(21) \citealt{sandell2005}, 
(22) \citealt{slysh1994}, 
(23) \citealt{tafalla1993}, 
(24) \citealt{valdettaro2002}, 
(25) \citealt{vanderwalt1995}, 
(26) \citealt{wilking1989}, 
(27) \citealt{wouterloot1993}, 
(28) \citealt{wouterlootbrand1989}, 
(29) \citealt{zhang2005}, 
(30) \citealt{zheng2001},
(31) \citealt{felli2007}.
\end{list}
}
\end{center}
\label{tsummary}
\end{table*}
%-------------------------------------------------------------------------------

\subsection{On the evolutionary stages of the sources of the survey \label{sdis}}

In order to estimate the evolutionary stage of the sources presented
in this paper, we searched the literature for the presence of dense
gas, outflow and maser emission. All this information is summarized in
Table~\ref{tsummary}.  The table shows that there are five sources
with dense gas and maser emission associated, being thus possibly the
youngest regions in the survey. This is supported by the evolutionary
classifications of \citet{hill2007} and \citet{longmore2007} for the
formation of high-mass stars.  According to these authors, the
earliest stages corresponding to a high-mass "starless core" are
characterized by cool ($\lesssim\!20$~K) millimeter/dense gas emission
with no maser or centimeter emission associated yet. After this stage,
the first signposts of the massive protostar are traced by warm
($\gtrsim\!50$~K) millimeter emission and class I/II-methanol maser
emission \citep{ellingsen2007}, which corresponds to the "hot-core"
phase, not detected in the centimeter range yet. Finally, when the YSO
starts the hydrogen burning, it starts ionizing the surrounding gas
and it can be detected in the centimeter range, while the YSO is still
accreting part of its final mass. H$_2$O and OH maser emission may be
present at this last stage \citep{ellingsen2007}. Thus, as a first
approach, the \uchii\ regions associated with millimeter continuum
emission, dense gas and maser emission can be considered younger than
the \uchii\ regions without.

In an attempt to better classify our \uchii\ regions in an evolutive
sequence, we studied the morphology of the centimeter and millimeter
emissions, and the relation between them. From the separation between
both peaks of emission, listed in Table~\ref{tsummary}, one can see
that there are four \uchii\ regions clearly associated with a compact
millimeter peak (I04579, I19045, I22198, and NGC\,7538-IRS9),
suggesting that the natal cloud has not yet been significantly
disrupted by the massive YSO. 

We quantified the degree of disruption of the natal cloud by comparing
the millimeter flux density encompassed in a region very close to the
centimeter peak position with the total millimeter flux density.  More
precisely, on one hand, the total millimeter flux density was
evaluated as the flux density inside a $0.5$~pc-diameter region
\citep[the typical size of massive dense cores, \eg][]{beltran2006b,
zinneckeryorke2007}, centered on the centimeter peak position,
$S_{0.5}$. The fixed diameter of 0.5~pc allows us to discard the
contamination of the measured flux density from nearby YSO or dense
clouds. On the other hand, we evaluated the millimeter flux density
inside a $0.2$~pc-diameter region (the size of the IRAM~30\,m beam at
the distance of the farthest source of this survey) centered on the
position of the centimeter peak, $S_{0.2}$. We defined the
\emph{disruption degree} as
%
%-------------------------------------------------------------------------------
\begin{equation}
{\rm Disruption~degree}
=
1-\frac{S_{0.2}}{S_{0.5}}\,
.
\label{eq_disruption_degree}
\end{equation}
%-------------------------------------------------------------------------------
%
Thus, a low disruption degree indicates that the millimeter emission
is concentrated inside a $0.2$~pc-diameter region around the
centimeter peak position, while a high disruption degree indicates
that the millimeter emission spreads out in a $0.5$~pc-diameter
region.  Taking into account the pointing errors of the IRAM~30\,m
telescope, we estimated an error of $\sim\,4\%$ for the values of the
disruption degree.

\begin{figure*}[ht!]
\begin{center}
\begin{tabular}{c}
\noalign{\bigskip}
	\epsfig{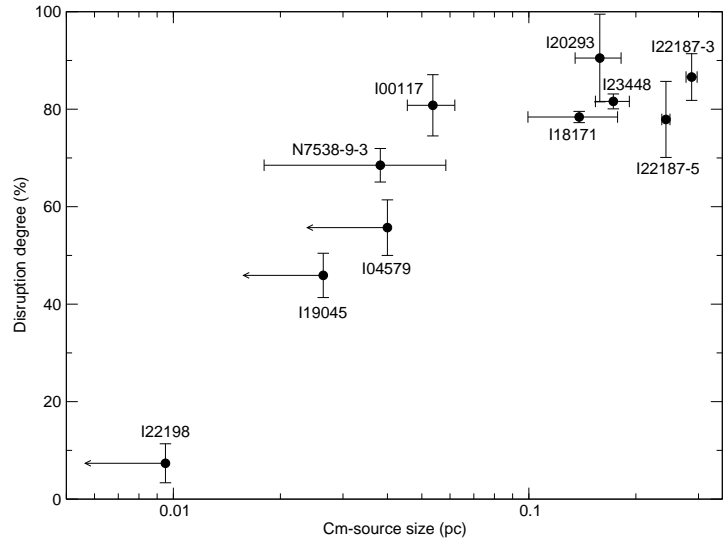} \\
\noalign{\bigskip}
\end{tabular}
\caption{{\small
Disruption degree of the natal cloud defined as one minus the ratio of the
millimeter flux density picked up within a $\sim\!0.2$~pc-diameter region (the
size of the IRAM~30\,m beam at the distance of the farthest source of this
survey) centered on the centimeter peak position and the millimeter flux density
picked up within a $\sim\!0.5$~pc-diameter region (the typical size of massive
dense cores) also centered on the centimeter peak position, versus the size of
the centimeter source (from Table~\ref{tphyspar}). For the disruption degree the
errors have been estimated taking into account the pointing errors of the
IRAM~30\,m telescope. For the cm-size we estimated the error from the gaussian
fit obtained with the task \texttt{JMFIT} in AIPS.
}}
\label{fimmcmplot}
\end{center}
\end{figure*}

The disruption degree defined in this way should evaluate to what
extent the massive YSO has pushed out and dispersed the natal gas and
dust. In Table~\ref{tsummary} the regions are ordered from smallest to
largest disruption degree, listed in column (6). From the table, one
can see that the sources with smallest disruption degrees, $<70\%$,
are I22198, I19045, I04579, and NGC\,7538-IRS9. If the cloud is
disrupted by the expansion of the \uchii\ region then one would expect
to find a correlation between the size of the centimeter source and
the disruption degree. Such a correlation is clear in
Fig.~\ref{fimmcmplot}, where we plot  the disruption degree versus the
size of the centimeter source.

%---------------------------------------------------------------------
%---------------------------------------------------------------------

It should be mentioned that the disruption degree and the centimeter
size could be affected not only by the expansion of the \hii\ region
(an evolutionary effect), but also by the initial density of the natal
cloud. For example, \cite{depree1995} find that for initial higher
densities of the natal cloud the sizes of the \uchii\ regions
associated are smaller. Similarly, the disruption degree could be
affected by the initial density of the cloud: the smaller the initial
density, the larger the disruption degree. In addition, inhomogenities
in the medium surrounding the massive YSO could produce an
accumulation of matter where the density is higher, shifting the peak
of millimeter emission from the centimeter peak, as could be the case
of I22187 (see \S~\ref{i22187}). However, several arguments are
against the interpretation of the disruption degree as tracing only
the initial conditions. First, in the first stages of the formation of
a massive YSO (assuming no close companions), the YSO has to be
associated with millimeter flux at least as strong as its surroundings
\citep[\eg][]{hill2007}. Second, we found small values of the
disruption degree for the regions which we previously classified as
less evolved judging from dense gas and maser emission (see above). In
addition, we note that our sample does not show an important
correlation between the size of the centimeter source and the
bolometric luminosity.  So, the disruption degree seems to be tracing
more the evolution of the massive YSO than the initial density of the
cloud.

Even more relevant is that our classification from the disruption
degree is consistent with the infrared properties and in particular
with the infrared excess derived from the 2MASS data, which should
also be indicative of the evolutionary stage (\S~\ref{sres2mass}). We
found that the source with the smallest disruption degree in our
survey, I22198, has no near- (2MASS source) neither mid-infrared (MSX
source) emission, suggesting that this YSO is deeply embedded in gas
and dust, which is opaque to the infrared emission. Regions with
(slightly) larger disruption degree are I19045, I04579, and
NGC\,7538-IRS9, the three showing red or orange infrared excess
(\S~\ref{sres2mass}). For these sources the centimeter emission is
compact and spatially coincident with the millimeter emission, which
is compact as well. Sources with larger disruption degrees as I18171
and I00117, for which the millimeter emission appears clumpy,
extended, and/or displaced from the centimeter peak, have smaller
infrared excesses (yellow).  Finally, sources with the largest
disruption degrees (I23448, I20293) have the centimeter emission less
compact, with the millimeter emission dispersed around the centimeter
source, and the infrared excess being among the smallest in the
sample.  Thus, our sample contains intermediate/high-mass star-forming
regions which can be ordered in an evolutionary sequence, from regions
containing massive YSOs deeply embedded in the natal cloud, to regions
with massive YSOs whose ionizing photons are in the process of
completely disrupting the natal cloud.

\section{Conclusions \label{scon}}

We observed, with the IRAM~30\,m telescope and the VLA, the continuum
emission at 1.2~mm, 1.3~cm and 3.6~cm toward a sample of 11
intermediate/high-mass star-forming regions not previously observed at
millimeter and/or centimeter wavelengths. Nine were observed in the
millimeter range and 10 in the centimeter range (8 coinciding with the
regions observed at IRAM~30\,m). Additionally, four of the most
promising regions were observed with the VLA at 7~mm. Our main
conclusions can be summarized as follows:

\begin{list}{}{}

\item[1.] From the IRAM~30\,m observations we found 1.2~mm emission in
all regions, but I18212. Four regions have very extended and weak
emission, while the other six show a clear emission peak with some
substructure. We derived the mass of gas and dust traced at 1.2~mm,
finding values from 10 to 140~\mo.

\item[2.] From the centimeter VLA observations we found 3.6~cm
emission in all regions, being most of them from compact sources. At
1.3~cm, all regions, but I18123, have emission associated with some
3.6~cm sources and a centimeter source is found within or near the
IRAS error ellipse. We estimated the spectral indices of the central
sources (those lying within the central $50\arcsec$ around the IRAS
source) between 3.6 and 1.3~cm. Additionaly, we searched the NVSS for
emission at 21~cm, and estimated the spectral index between 21 and
3.6~cm. For most of the sources associated with the IRAS source we
found flat or positive spectral indices, suggesting that the emission
comes from compact or UC \hii\ regions, thermal radiojets or
externally ionized globules. Deconvolved sizes of the centimeter
sources range from <0.01 to 0.3~pc, and the physical parameters of the
\hii\ regions indicate that the ionizing stars are early B-type stars.

\item[3.] For the four regions observed at 7~mm the emission was
partially resolved. Taking into account the spectral index between 3.6
and 1.3~cm, we were able to derive the dust contribution to the 7~mm
continuum emission for I04579 and I22198, which is of about 38\% and
44\%, respectively.

\item[4.] By combining our data with the infrared surveys of 2MASS,
MSX, and IRAS, and with IRAC-\emph{Spitzer} data (when available), we
built the SED and fitted a modified blackbody law taking into account
the contribution of a free-free optically thin law for the centimeter
observations. We found median values of $\sim\!25$~\mo\ for the mass,
$\sim\!29$~K for the dust temperature, and $\sim\!1.9$ for the dust
emissivity index for the envelope of each intermediate/high-mass YSO.

\item[5.] By studying the morphology of the centimeter and millimeter
emission, and the separation between both peaks of emission, we found
a correlation between the degree of disruption of the natal cloud
(estimated from the millimeter flux picked up within a
$\sim\!0.2$~pc-diameter region, centered at the position of the
centimeter peak, and the millimeter flux picked up within a
$\sim\!0.5$~pc-diameter region), and the size of the centimeter
source. This allowed us to establish an evolutionary sequence for the
\uchii\ regions studied in this work, which is consistent with the
expected evolutionary stage from the dense gas, outflow, and maser
emission, as well as with the infrared properties of the sources
associated with the \uchii\ regions. Thus, we classified our sources
from the youngest, I22198, similar to the Class 0 low-mass sources,
with no near- neither mid-infrared emission, to the most evolved,
I23448 and I20293, with the centimeter source(s) having disrupted the
molecular natal cloud.

\end{list}

\begin{acknowledgements}
\'A.\ S. is grateful to G. Sandell for providing us the 1.3~mm image
of NGC\,7538-IRS9. We are grateful to Stan Kurtz, the referee, for his
valuable comments and suggestions, which have helped to improve the
scientific content of this paper. The authors are supported by a MEC
grant AYA2005-08523-C03 and FEDER funds. This publication makes use of
data products from the Two Micron All Sky Survey, which is a joint
project of the University of Massachusetts and the Infrared Processing
and Analysis Center/California Institute of Technology, funded by the
National Aeronautics and Space Administration (NASA) and the National
Science Foundation. Furthermore, this publication benefits from data
of the Midcourse Space Experiment. Processing of the data was funded
by the Balistic Missile Defense Organization with additional support
from NASA Office of Space Science.
\end{acknowledgements}

\end{document}